\documentclass[twocolumn,bibyear]{aa}  
\usepackage{natbib}
\usepackage{graphicx}
\usepackage{txfonts}
\usepackage{amsmath}
\usepackage{amssymb}
\usepackage{xcolor}
\definecolor{lightbrown}{rgb}{0.71, 0.40, 0.11}
\definecolor{darkblue}{rgb}{0.0, 0.0, 0.55}
\definecolor{darkorange}{rgb}{1.0, 0.55, 0.0}
\usepackage{hyperref}
\hypersetup{linkcolor=blue,citecolor=blue,filecolor=cyan,urlcolor=cyan}
\usepackage{colortbl}

\definecolor{mycyan}{HTML}{00E5E5}
\definecolor{mytan}{HTML}{D2B48C}
\definecolor{myblue}{HTML}{00008B}
\definecolor{myorange}{HTML}{FF8C00}
\definecolor{mypink}{HTML}{FFC0CB}

\begin{document} 

   \title{Clustering Wind data at 1 AU to contextualize magnetic reconnection in the solar wind}
   \author{Francesco Carella\inst{1}
            \and Giovanni Lapenta\inst{1}\textsuperscript{\textdagger}
            \and Alessandro Bemporad\inst{2}
            \and Stefan Eriksson\inst{3}
            \and Maria Elena Innocenti\inst{4}
            \and Sophia Köhne\inst{4}
            \and Jasmina Magdalenic\inst{1,5}
            }

   \institute{KU Leuven, Department of Mathematics, Celestijnenlaan 200b, Leuven, 3001, Belgium\\ \email{francesco.carella@kuleuven.be}
            \and National Institute for Astrophysics, Astrophysical Observatory of Turin, Via Osservatorio 20, Pino Torinese, 10025, Turin, Italy
            \and Laboratory for Atmospheric and Space Physics, University of Colorado, Boulder, CO, USA
            \and Institut für Theoretische Physik, Ruhr-Universität Bochum, Universitätstraße 150, 44801, Bochum, Germany
            \and Royal Observatory of Belgium, Avenue Circulaire 3, 1180, Brussels, Belgium
            }
          
      \date{\textsuperscript{\textdagger}Died in May 28, 2024.\\[0.2cm]
      Received: 26 February 2025; Accepted: 25 July 2025}

   \abstract  
   {Magnetic reconnection events are frequently observed in the solar wind. Understanding the patterns and structures within the solar wind is crucial to put observed magnetic reconnection events into context, since their occurrence rate and properties are likely influenced by solar wind conditions.}
   {We employed unsupervised learning techniques such as self-organizing maps (SOM) and K-Means to cluster and interpret solar wind data at 1 AU for an improved understanding of the conditions that lead to magnetic reconnection in the solar wind.}
   {We collected magnetic field data and proton density, proton temperature, and solar wind speed measurements taken by the Wind spacecraft. After preprocessing the data, we trained a SOM to visualize the high-dimensional data in a lower-dimensional space and applied K-Means clustering to identify distinct clusters within the solar wind data.}
   {Our analysis revealed that the reconnection events are distributed across five different clusters: a) slow solar wind, b) compressed slow wind, c) highly  Alfvénic wind, d) compressed fast wind, and e) ejecta. Compressed slow and fast wind and ejecta are clusters associated with solar wind transients such as stream interaction regions and interplanetary coronal mass ejections. The majority of the reconnection events are associated with the slow solar wind, followed by the highly Alfvénic wind, compressed slow wind, and compressed fast wind, and a small fraction of the reconnection events are associated with ejecta.}
   {Unsupervised learning approaches with SOM and K-Means lead to physically interpretable solar wind clusters based on their transients and allow for the contextualization of magnetic reconnection exhausts' occurrence in the solar wind.}


   \maketitle
%
%-------------------------------------------------------------------
\section{Introduction} \label{sec:intro}
\begin{figure*}[!t]
	\centering
    \includegraphics[width=0.93\linewidth]{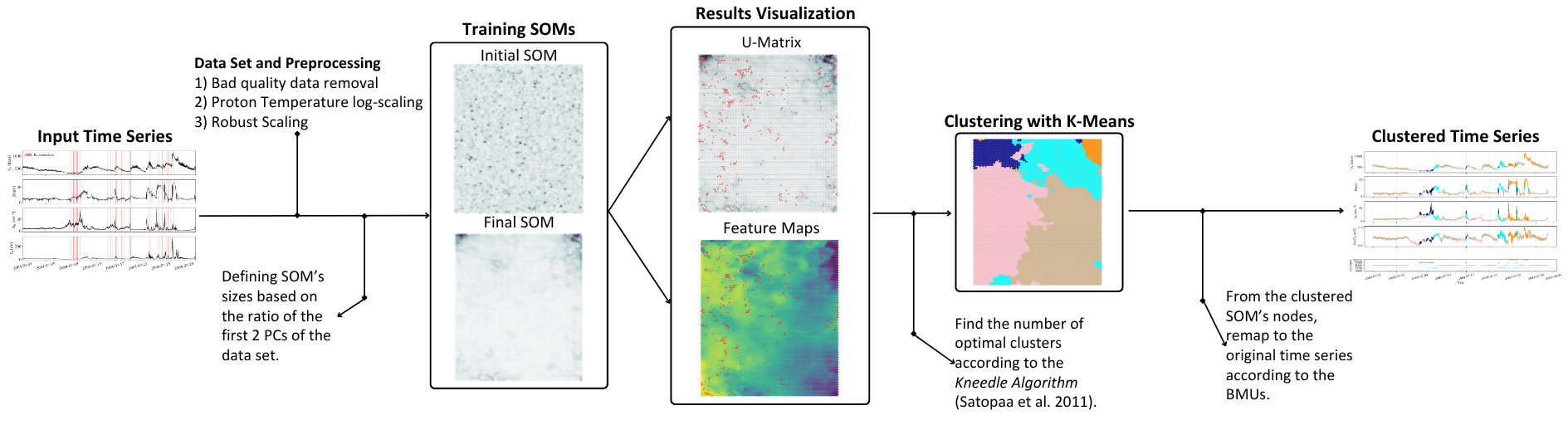}
	\caption{Schematic representation of the workflow. From left to right: The dataset was preprocessed by removing low-quality data and applying robust scaling, and then the SOM was trained on the preprocessed dataset. After this, the K-Means algorithm was applied to the SOM's nodes to identify distinct clusters within the solar wind data. Solar wind data series are colored according to the cluster we identify. }
	\label{fig:flow}
\end{figure*}

The Sun constantly emits into interplanetary space the solar wind, a stream of magnetized particles mostly consisting of protons, electrons, and alpha particles.
The concept of the solar wind was first introduced by Eddington in 1910 \citep{eddy1910} as an explanation for the observed shapes of cometary tails, and its origins were clarified by Parker in 1958 based on theoretical arguments~\citep{Parker}. First observations by the Luna-1 and Mariner-2 missions quickly followed~\citep{gringauz1960study, neugebauer1962solar}. Since then a large number of observations and theoretical studies have been made and several missions such as Ulysses \citep{WENZEL198925}, Wind \citep{ogilvie1997}, ACE (Advanced Composition Explorer) \citep{stone98}, and more recently Parker Solar Probe (PSP) \citep{Fox2016} and Solar Orbiter (SolO) \citep{Muller2020} have been launched to probe the solar wind at different distances from the Sun.\\
The solar wind has traditionally been classified according to the average speed of protons measured at 1 AU \citep{sw} into fast and slow solar wind. The fast wind (\(v \gtrsim 500\) km/s) generally originates from the coronal holes \citep{bd1971,Cranmer2009-mu}, i.e., transient coronal regions characterized by less dense and cool plasma, from which the Sun's magnetic field extends into interplanetary space. The slow solar wind (\(v \lesssim 400\) km/s) has many more variable plasma properties \citep{Schwenn2006-uq} than the fast wind and different candidate solar sources \citep{Abbo2016-cn}.\\
Recent studies have shown that it is necessary to go beyond the bimodal solar wind model. For instance, Alfvénicity is a characteristic generally attributed to fast wind of coronal hole origin, but recent work has identified highly Alfvénic slow solar wind parcels \citep{DAmicis_2015}, and recent PSP observations have shown measurements of generally slow, highly Alfvénic solar wind (\citet{Bale2019, Kasper2019}).\\
Several investigations have tried to classify solar wind by taking into account different aspects, such as the solar origin. \citet{Xu} proposed a four-mode plasma categorization scheme that distinguishes between ejecta, wind of coronal hole origin, sector reversal origin, and streamer belts origin based on three parameters. This classification scheme is the starting point for \citet{Camporeale}, who added a few additional parameters from a supervised learning approach based on a Gaussian process (GP) for the classification.\\
At 1 AU we measure both pristine solar wind and the effects of interplanetary transients such as interplanetary coronal mass ejections (ICMEs) and stream interaction regions (SIRs). The ICMEs \citep{Kilpua2017,Gopalswamy_2016} are macro-scale structures that originate as magnetized plasma clouds ejected from the Sun and are also known as simply coronal mass ejections (CMEs). The SIRs are large-scale plasma structures that result in regions of compression as a fast solar wind overtakes a slow solar wind. A pair of forward and reverse shocks can be produced from this interaction. If the coronal hole source region of the fast wind survives one or more solar rotations, we refer to these SIRs as corotating interaction regions (CIRs) \citep{heber}. These transients and high-speed solar wind streams can result in significant space weather impacts when they interact with the Earth's magnetosphere \citep{2018SSRv..214...17B}. ICMEs may have particularly strong space weather effects, whereas fast streams and CIRs produce more modest space weather impacts \citep{ward}.\\
The solar wind is also affected by smaller-scale phenomena, such as magnetic reconnection \citep{Gosling2012}. Although ICMEs are caused by magnetic reconnection at the solar surface and corona, their passage through interplanetary space causes magnetic reconnection at larger distances from the Sun, in the solar wind \citep{Chian_2011}.\\
\citet{eriksson2022} have studied the characteristics of magnetic reconnection in multi-scale current sheets in the solar wind.  This provided a method of recognizing the occurrence of magnetic reconnection in solar wind in situ observations. With this method, \citet{eriksson2022} produced a dataset of magnetic reconnection exhausts that took place between July 1, 2004, and December 31, 2014, counting a total of 3374 reconnection exhausts.
\cite{HEIDRICHMEISNER2018397} used K-Means clustering to classify solar wind data from the ACE spacecraft dataset. They showed the importance of the combination of the magnetic field strength and the lower-order moments of the proton velocity distribution function, such as proton density, temperature, and speed, while excluding any ICMEs in their dataset classification.
In this work, we apply unsupervised learning techniques to discover patterns in in situ observations of the solar wind at 1 AU, with the objective of finding correspondences between the occurrence of magnetic reconnection and solar wind transients, including ICME transients. The workflow is shown in Figure \ref{fig:flow} and it consists of the following steps:
\begin{enumerate}
   \item \textit{Data collection:} Retrieve magnetic field data from the MFI instrument \citep{1995} and proton density, proton temperature, and solar wind speed data from the 3DP instrument \citep{3dp} on board the Wind spacecraft \citep{ogilvie1997}, focusing on one month of data. This step is described in Section \ref{sec:data}.
   \item \textit{Preprocessing:} Clean the dataset by removing low-quality data points and applying feature scaling, ensuring that the data are on a comparable scale. This activity is described in Section \ref{sec:preprocessing}.
   \item \textit{Training SOM:} Utilize the self-organizing map (SOM) \citep{som1} to transform time series data into a visual map. This is a technique that has already been used to cluster ACE data \citep{amaya2020visualizing}, magnetospheric simulations \citep{innocenti21}, and observations \citep{Edmond2024}, and fully kinetic simulations of tearing and/or plasmoid instabilities~\citep{kohne2023unsupervised}. The SOM helps visualize high-dimensional data in a lower-dimensional space, making it easier to understand the underlying patterns and structure of the solar wind data. We described SOM training in Section \ref{sec:som}.
   \item \textit{Clustering:} Apply the K-Means \citep{lloyd,macqueen1967some} clustering algorithm to the trained SOM to identify distinct clusters within the solar wind data and determine the optimal number of clusters using the Kneedle algorithm. We describe these steps in Section \ref{sec:kmeans}.
   \item \textit{Analysis:} Use the reconnection exhausts catalog from~\citet{eriksson2022} to identify the occurrences of magnetic reconnection within the solar wind data and examine the distribution of magnetic reconnection data across the identified clusters. The aim is to understand the relationship of reconnection events with different solar wind conditions. Then, we compare the results with the \citep{Xu} classification of the solar wind. We describe this in Section \ref{sec:results}.
\end{enumerate}
\section{Dataset} \label{sec:data}
\begin{figure}[!hbt]
	\centering
	\includegraphics[width=0.95\linewidth]{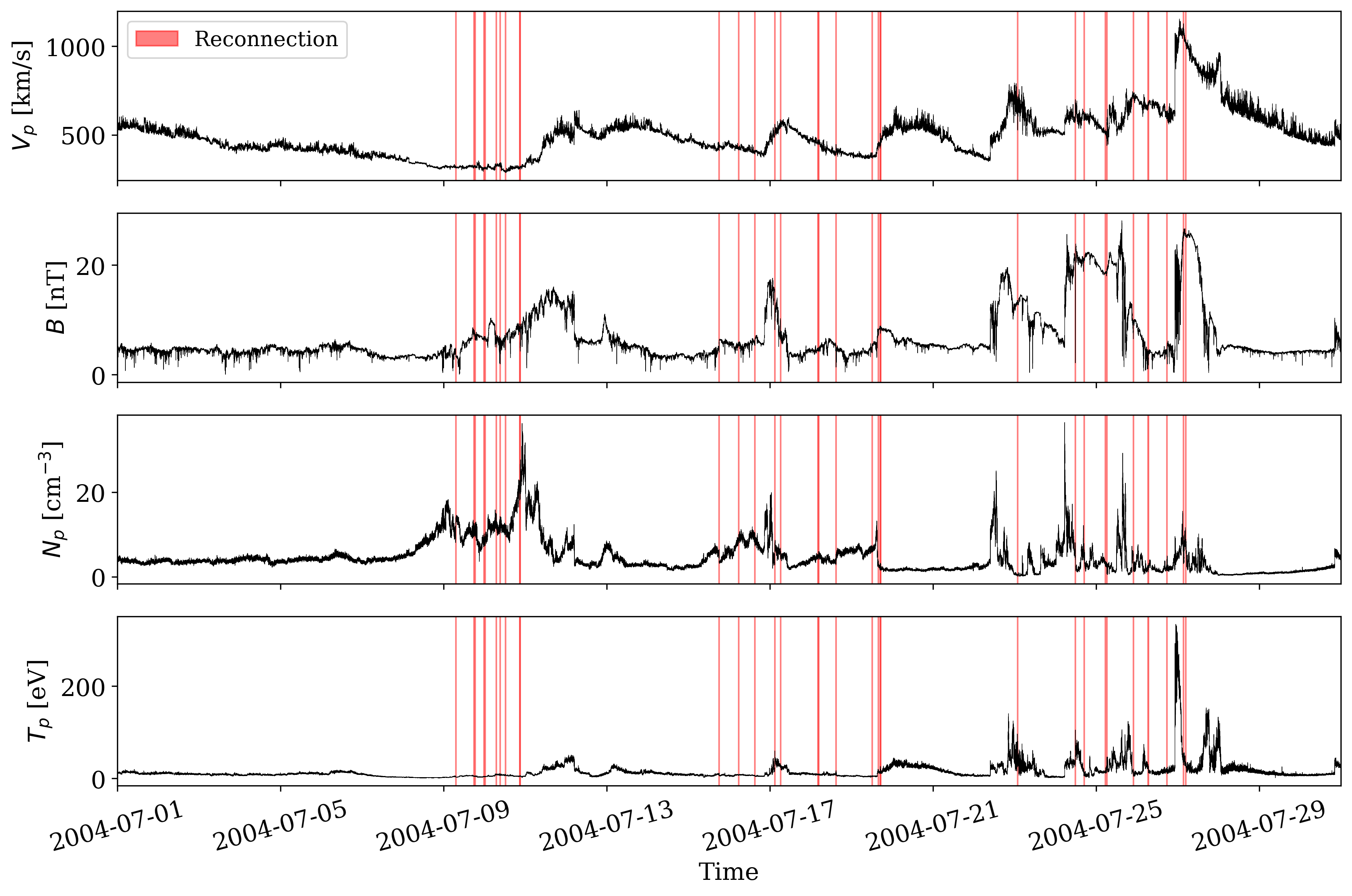}
	\caption{ Proton speed, \(V_p\), magnetic field strength, \(B\), proton density, \(N_p\), and proton temperature, \(T_p\), time series for July 2004. In red are the reconnection exhausts.}
	\label{fig:bvmag04}
\end{figure}
We used data from the Wind spacecraft \citep{ogilvie1997}; in particular, magnetic field data from the MFI instrument \citep{1995} and velocity, proton density, and proton temperature data from the 3DP instrument \citep{3dp}.
Data were retrieved from the CDAWeb\footnote{Coordinated Data Analysis Web \url{https://cdaweb.gsfc.nasa.gov/}} Wind dataset using the SunPy Python library \citep{sunpy_community2020}. The reconnection exhausts' catalog in \citet{eriksson2022} covers the period from 2004 to 2014. In our case, we chose the last six months of 2004 (from July 1, 2004, 00:00:00 to December 31, 2004, 23:59:59), corresponding to the descending phase of the XXIII Solar Cycle.
We focused on four different physical quantities, with a 3 s cadence: the magnetic field strength, \(B\), measured by the MFI instrument, the proton temperature, \(T_p\), the proton density, \(N_p\), and the proton speed, \(V_p\), measured by the 3DP instrument.
The initial dataset consists of more than 5 million data points for a total of 181 reconnection exhausts (7556 points). Figure \ref{fig:bvmag04} shows the month of July of the initial dataset as an example, with the four features plotted as time series. In red, we highlight the reconnection exhausts from the \citep{eriksson2022} catalog. 

\subsection{Data cleaning} \label{sec:preprocessing}
The Wind data was cleaned by filtering the data gaps and the low-quality data using the data flags from \textit{WI\_PM\_3DP} database on the CDAWeb; respectively, “GAP” to remove the gaps (0 is no gap and 1 is a gap) and “VALID” (1 is good-quality data and 0 is low-quality data) to remove the low-quality data. Low-quality data from the MFI instrument was also removed by filtering the 1e$^{-31}$ values. Whenever there was a data gap or low-quality data in at least one of the parameters, the entire set of corresponding data was removed.\\
After a visual inspection of the dataset, we noticed that some low-quality data had not been flagged, and the following time intervals had been manually removed: “September 10, 2004, 12:15-13:00,” “October 23, 2004, 04:19-04:20,” and “December 17, 2004, 17:43-17:44.” All of those time intervals were characterized by sudden nonphysical spikes in one or more parameters (see Figure \ref{fig:badata}).

\subsection{Features scaling}
As is shown in Figure \ref{fig:bvmag04}, the proton temperature is characterized by a large dynamic range, which can negatively impact the performance of machine learning algorithms. Therefore, we decided to use the logarithm of the proton temperature for our clustering. We remark that the proton temperature is fundamental in distinguishing fast solar wind, which is generally hotter, from slow solar wind, which is generally colder (\citet{marsch_solar_1982, cranmer_coronal_2002}). Furthermore, spikes in the proton temperature can be associated with shocks.\\
Since machine learning algorithms are sensitive to the scale of the input features (\citet{huang, fscaling}), the second step of the preprocessing phase was the scaling of the features. Scaling of a dataset is a common requirement for many distance-based machine learning estimators. Outliers, i.e., data points significantly out of range with respect to the other points in the dataset, may have a negative impact on the mean and the variance of the dataset.
Following  \citet{kohne2023unsupervised}, we used robust scaling to avoid performance reduction and scaling biases due to outliers. Robust scaling \citep{Rousseeuw} consists of removing the median and scaling the data according to the interquartile range, which in our case is between the first and third quartile.
\section{Methodology} \label{sec:floats}
\subsection{Self-organizing maps}
\label{sec:som}
Self-organizing maps \citep{som1} are neural networks that are used in clustering and data visualization, with the aim of producing an ordered and low-dimensional representation of a higher-dimensional dataset. The low-dimensional space generally consists of a structured 2D grid composed of \(x_{dim} \times y_{dim}=K\) nodes (also called units or neurons). Here, \(x_{dim}\) represents the number of columns and \(y_{dim}\) the number of rows. Each neuron within this grid possesses a unique position and is linked to a weight vector, \(\Vec{w}_i \in \mathbb{R}^m\), where \(m\) is the number of features considered.\\
To train the SOM, the matrix of weights is initialized. This initialization can be done either randomly or by spanning the weights along the 2D linear subspace of the first two principal components. Then, each data point, \(\Vec{x}\), is randomly picked from the dataset and presented to the network. For each iteration, the following procedure (competition, collaboration, and adaptation) is repeated:
\begin{enumerate}
	\item \textit{Competition (or similarity matching)}\\
	The SOM nodes competed among themselves to become activated: if \(\mathbb{W}\) is the set of all the weights in the map, the best matching unit (BMU) of the input data point, \(\Vec{x}\), is identified, by choosing the node whose weight, \(\Vec{w}_b \in \mathbb{W}\), has a minimum distance (\(||\cdot||\)) with respect to \(\Vec{x}\) of
	\begin{equation}
		\Vec{w}_b = \underset{\Vec{w}_i\in \mathbb{W}}{\text{argmin}}(||\Vec{x}-\Vec{w}_i||).
		\label{eq:bmu}
	\end{equation}
	The BMU is the node that best represents the input data point.\\
   The distance between the input data point and the weights can be calculated using different distance metrics, such as the Chebyshev distance, the Euclidean distance, the Mahalanobis distance, the Tanimoto distance, the Manhattan distance, or the vector dot product. In this work, as is suggested in \citet{som4}, we opted for the Euclidean distance in order to be consistent with the distance metric that was adopted in the K-Means step.
   
\item \textit{Collaboration}\\
The identified BMU and its neighboring units in the SOM are activated through a lattice neighborhood function, \(h\left(\sigma(\tau),i,b\right)\) (Eq. \ref{eq:nfun}). Here \(\sigma(\tau)\) represents the iteration-dependent “lattice neighborhood width” (or “neighborhood radius”), usually a monotonically decreasing function that effectively determines the extension of the neighborhood during the training, \(i\) is the index of the i-th neuron and \(b\) is the BMU index.\\
To obtain a smooth neighborhood function, a Gaussian function is usually selected:
\begin{equation}
	h(\sigma(\tau),i,b) = e^{-\frac{||\Vec{w}_i-\Vec{w}_b||^2}{2 \sigma(\tau)^2}}.
    \label{eq:nfun}
\end{equation}
\item \textit{Adaptation}\\
For each iteration, the BMU and its neighbors are updated to increase the similarity to the presented sample, $\Vec{x}$, of the input space.\\
Each weight is updated according to the following rule:
\begin{equation}
    \Vec{w}_i(\tau+1) = \Vec{w}_i(\tau) + \eta(\tau)	h(\sigma(\tau),i,b) (\Vec{x}(\tau)-\Vec{w}_i(\tau))
    \label{eq:wupd}
,\end{equation}
where \(i\) represents the index of the node and \(b\) the BMU index. The magnitude of the update depends on the distance of the node, \(i\), from the BMU.  Here \(\eta(\tau) \in [0;1]\) is the “learning rate,” which is a decreasing function of the iteration step, \(\tau\).\\
The update of the weights depends on the learning rate, \(\eta(\tau)\), which controls the magnitude of the update, on the neighborhood function, \(h(\sigma(\tau),i,b)\), which controls which and how many weights should be updated, and on the difference, \(||\Vec{x}(\tau)-\Vec{w}_i||\), which shifts the weights toward the input data.\\
\end{enumerate}
The overall concept is that at the beginning of the training, since the distance between the nodes (weights) and input data is high, i.e., the similarity is low, large updates are needed and a wider number of weights are involved. After several iterations, when the map topology has already partially adapted to that of the input data, minor weight updates are needed and therefore a smaller number of weights is involved in each update, meaning that the updates become more targeted toward improving the numerical similarity of the weight values.\\
This similarity relation can be expressed in terms of the “quantization error,” \(Q_E\):
\begin{equation}
    Q_E = \frac{1}{n}\sum_{i=1}^n||\Vec{x}_i-\Vec{w}_{b|\Vec{x}_i}||^2
,\end{equation}
where \(n\) is the number of points in the dataset, \(\Vec{x}_i\) the i-th input data, and \(\Vec{w}_{b|\Vec{x}_i}\) the corresponding BMU given \(\Vec{x}_i\). It measures the average distance between each of the \(n\) entry data points and their BMU, and therefore how similar the map is compared to the input data distribution.\\
In this work, we used a CUDA implementation of SOM (CUDA-SOM, \citet{mistri2020cuda-som}) to speed up the learning process. The discussion and the scaling regarding this tool have already been examined in \citet{kohne2023unsupervised}.\\
The CUDA-SOM tool requires the following set of parameters as inputs:
\begin{itemize}
	\item Number of rows, \(x_{dim}\), and columns, \(y_{dim}\).
	\item The initial learning rate, \(\eta_0\), and the initial neighborhood radius, \(\sigma_0\).
	\item Type of decay (exponential or asymptotic) for learning rate and neighborhood radius.
	\item Distance metric to use. The options are Euclidean, sum of squares, Manhattan, and Tanimoto.
	\item Neighborhood function: Gaussian, Bubble, or Mexican hat.
	\item Type of lattice: square or hexagonal.
	\item Maximum number of epochs: \(T_{max}\).
    \item Learning mode: online or batch.
\end{itemize}
The parameters we chose are shown in Table \ref{tab:SOMet}. We explain in Appendix \ref{appendix:A} and Appendix \ref{appendix:B} the reasons for our parameter choices.\\
\begin{table}[!hbt]
	\centering
	\caption{CUDA-SOM settings.}
	\label{tab:SOMet}
   \resizebox{\linewidth}{!}{
	\begin{tabular}{|c|c|}
	\hline
	\textbf{Parameter} & \textbf{Value/Setup} \\ \hline
		Learning Mode & Online                 \\ \hline
		Number of Rows ($x_{dim}$) & 126                      \\ \hline
		Number of Columns ($y_{dim}$)              & 96                  \\ \hline
		Initial Learning Rate ($\eta_0$)              & 0.25                      \\ \hline
		Initial Neighborhood Radius ($\sigma_0$)      & 25                   \\ \hline
		Maximum Number of Epochs ($T_{max}$)              & 9                       \\ \hline
		Distance Metric              & Euclidean                       \\ \hline
		Neighborhood Function ($h(\sigma, i, b)$)              & Gaussian                      \\ \hline
		Initialization Method              & Random Input Vectors                       \\ \hline
		Lattice Type              & Hexagonal                        \\ \hline
		Grid Initialization Randomization              & On                       \\ \hline
		Decay Type for $\eta$ and $\sigma$              & Exponential                     \\ \hline
		Normalize Distance              & Off                     \\ \hline
	\end{tabular}
   }
\end{table}

\subsection{Clustering with K-Means}
\label{sec:kmeans}
After the SOM training, we used K-Means on the SOM nodes to obtain a lower number of clusters, which can be used to cluster solar wind time series. The K-Means algorithm \citep{lloyd,macqueen1967some} is a partitioning method that aims to minimize the intra-cluster variance and maximize the inter-cluster variance. The algorithm requires as an input the number of clusters, \(K\).\\
The number of clusters is a crucial parameter, and it can be estimated using the \textit{Kneedle} algorithm \citep{satopaa}. The algorithm is based on the computation of the second derivative of the intra-cluster variance, \(\text{ICV}(K)\), with respect to \(K\). The optimal number of clusters is the one that maximizes the second derivative of \(\text{ICV}(K)\).\\
The intra-cluster variance is a measure of the homogeneity of the clusters, and it is defined as
\begin{equation}
    \text{ICV}(K) = \sum_{k=1}^K \sum_{\Vec{x} \in C_k} ||\Vec{x}-\Vec{\mu}_k||^2
,\end{equation}
where \(C_k\) is the k-th cluster and \(\Vec{\mu}_k\) is the centroid of the k-th cluster.\\
To train the K-Means algorithm, we used the \textit{KMeans} class from the \textit{sklearn} library \citep{scikit-learn} with “k-means++” initialization, which is a method that chooses the initial centroids in a way that speeds up the convergence of the algorithm.\\
The results of the clustering of the SOM's nodes with K-Means are shown in Section \ref{sec:clustering}, and the Kneedle method results are shown in Appendix \ref{appendix:B}.

\section{Results}
\label{sec:results}
\begin{figure*}[!t]
    \centering
    \makebox[\textwidth][c]{
        \begin{tabular}{@{}c@{\hspace{0.5cm}}c@{}}
            \raisebox{-0.5\height}{\includegraphics[height=0.23\textheight]{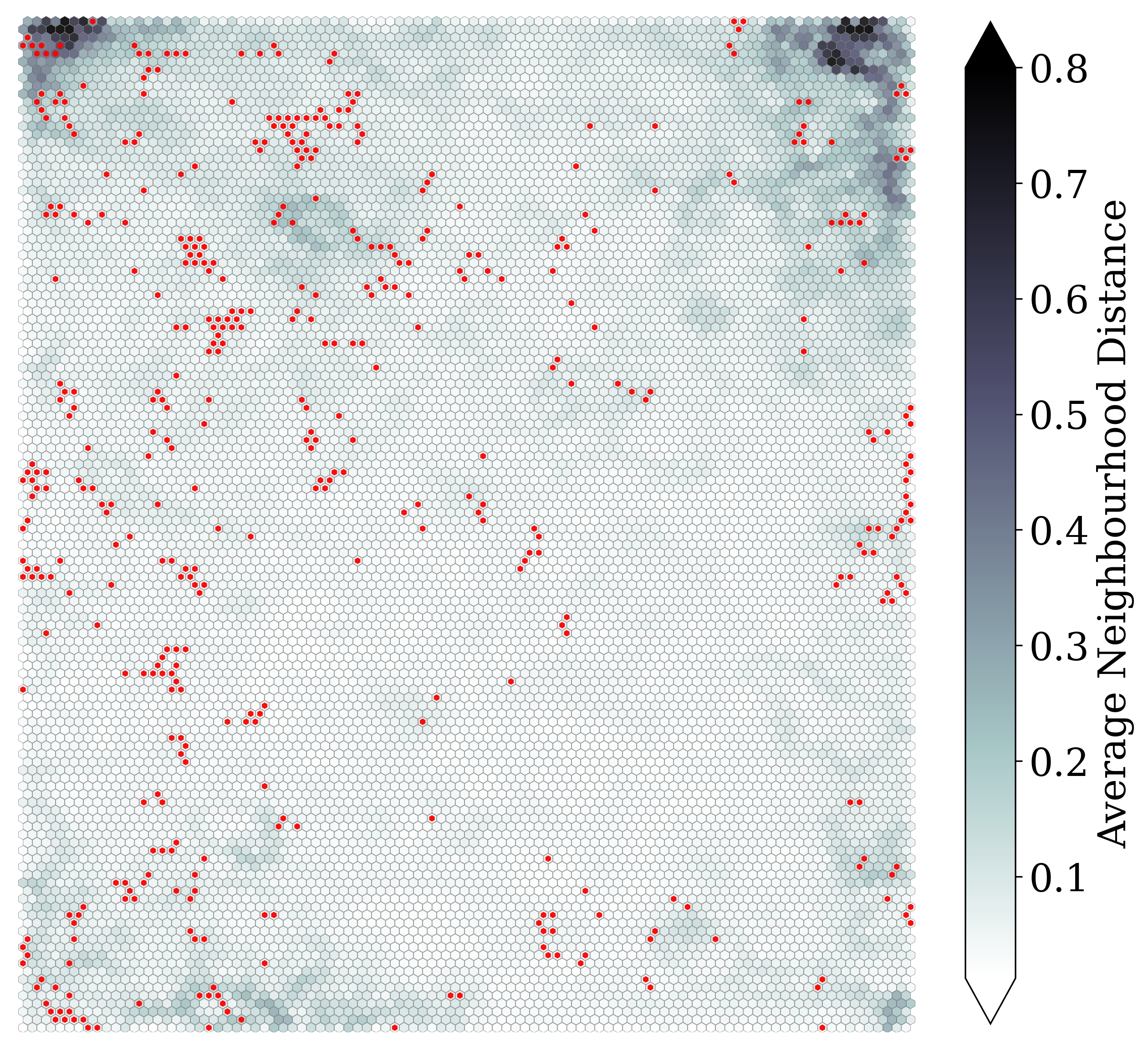}} &
            \begin{tabular}{cc}
                \includegraphics[width=0.23\textwidth]{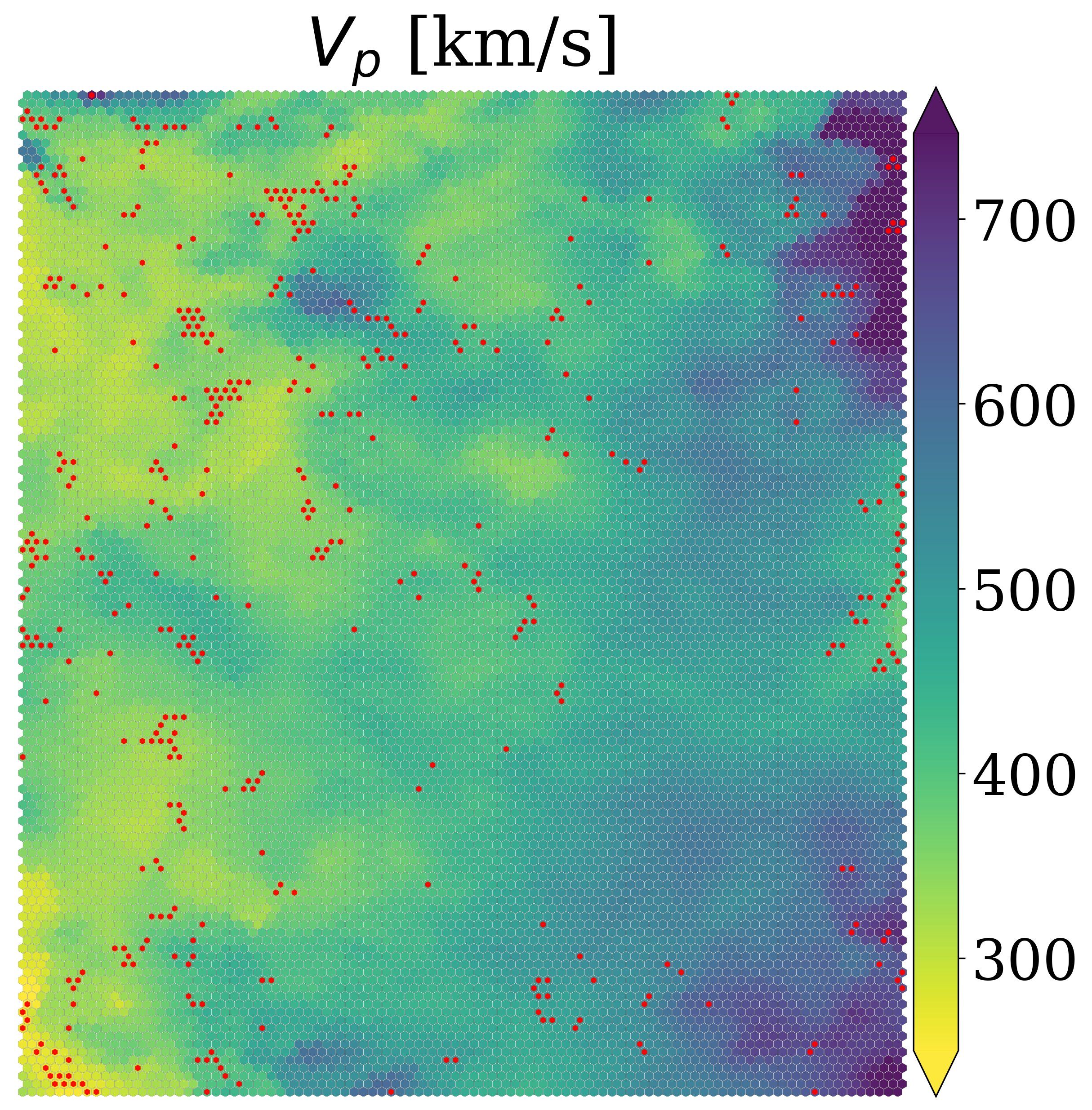} &
                \includegraphics[width=0.22\textwidth]{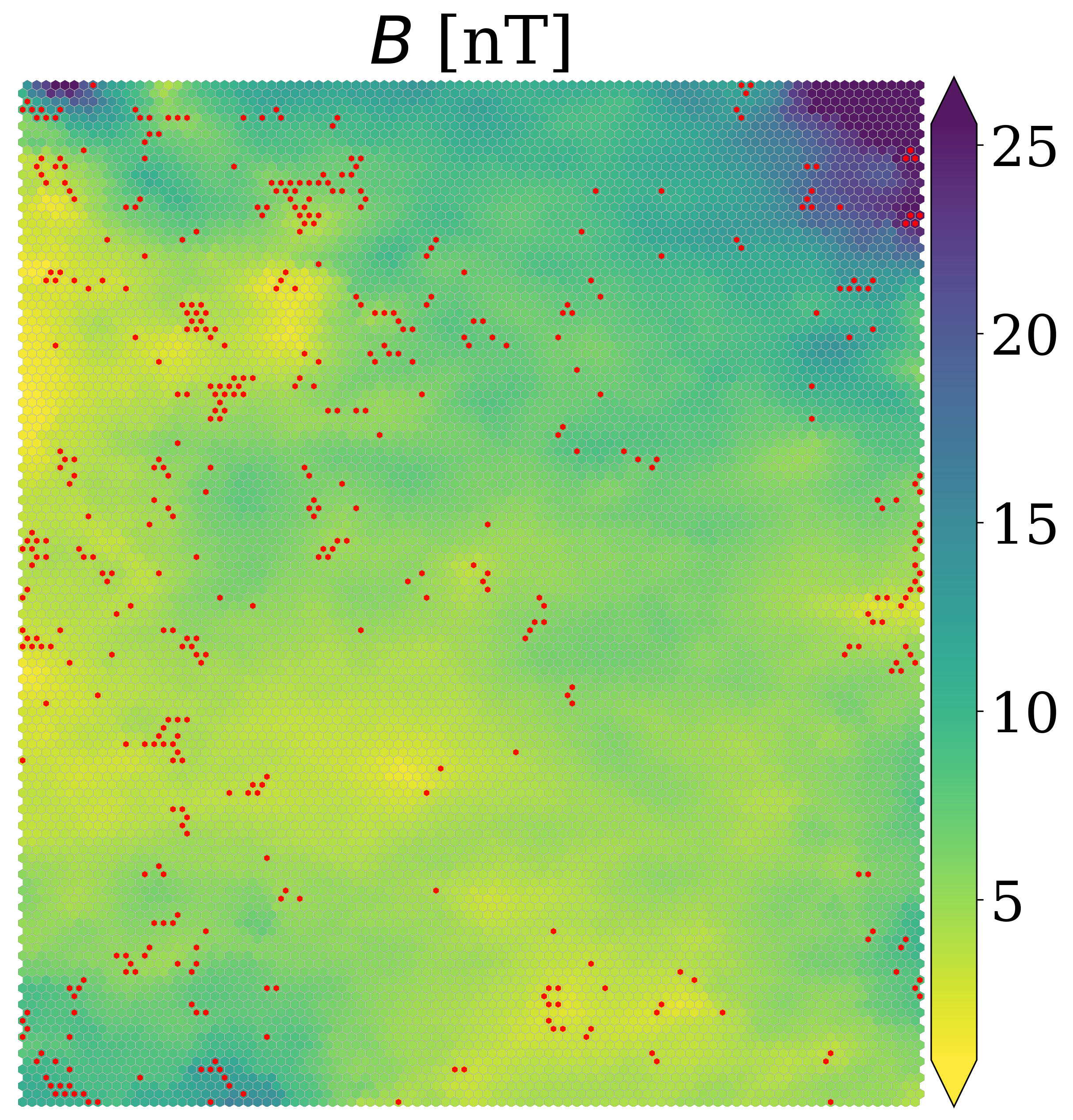} \\
                \includegraphics[width=0.22\textwidth]{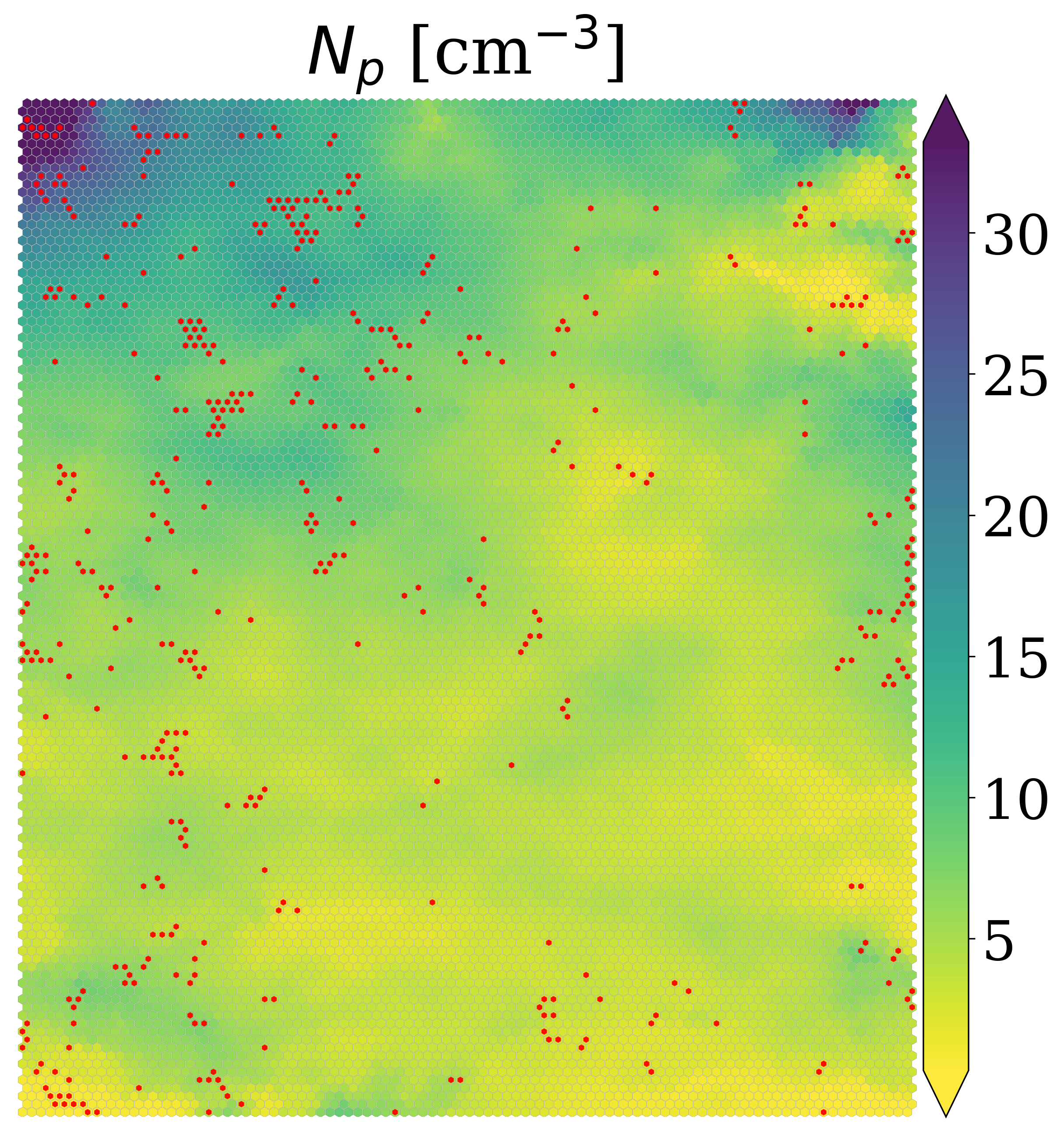} &
                \includegraphics[width=0.23\textwidth]{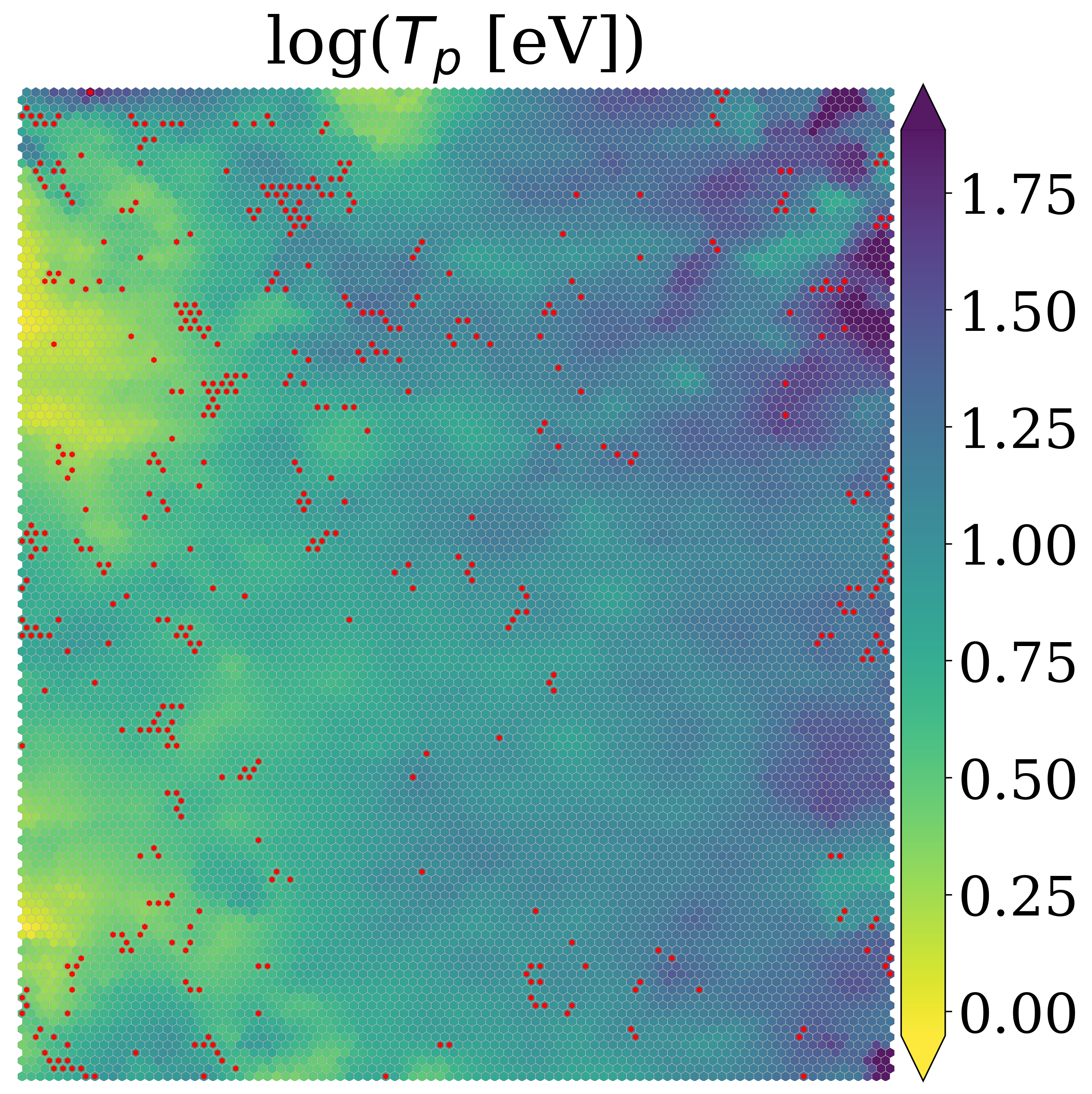} \\
            \end{tabular}
        \end{tabular}
    }
    \caption{(\textit{Left}) U-matrix representation of the SOM. (\textit{Right}) Feature maps: proton speed, magnetic field strength, proton density, and logarithm of the proton temperature. In red, we highlight the “reconnection neurons,” i.e., those with which more than four reconnection data points are associated. Here, the SOM weights have been rescaled according to the original dataset.}
    \label{fig:swhexmap}
\end{figure*}
\subsection{Visualizing the SOM results}
The training of the SOM was done on an NVIDIA L40S and took approximately 1.5 hours. The quantization error associated with the trained map is $Q_E = 0.11$. The trained SOM is shown in Figure \ref{fig:swhexmap}; each node in the map is represented by a hexagon. We show the U-matrix on the left and the feature maps corresponding to the four features on the right. The U-matrix represents the distance between the weights of the neurons. It is computed as the average distance between a neuron and its neighbors: a low value of the U-matrix corresponds to a region of the map where neurons are “more similar” to each other, while a high value of the U-matrix corresponds to a region of the map where neurons are “less similar” and more distant from each other.
This information can already be used to infer possible clusters in the dataset: low-value regions are likely within clusters, while high values are likely to be the boundaries between clusters.\\
We label the so-called ``reconnection neurons" in red. Each neuron represents multiple data points of the original dataset: since each data point has been matched with the reconnection exhausts data from \citet{eriksson2022}, it is possible to mark which neuron is the BMU for at least four reconnection data points. We chose four reconnection data points because they correspond to 12s, which is the median value of the exhaust time duration, shown in \citep{eriksson2022}.\\
On the right, we show the feature maps, representing the distribution of the features on the map. Here, each hexagon represents a neuron in the SOM, and the color indicates the value of the feature for that neuron. Since the original values for each neuron correspond to the scaled features, to visualize the actual values, the weights have been inversely scaled. We observe that the reconnection data are distributed in different regions of the map, characterized by a different combination of features (i.e., high magnetic field strength, low density, high velocity, medium temperature, and so on), suggesting that the reconnection data belong to different clusters.
\subsection{Clustering results}
\label{sec:clustering}
\begin{figure*}[!hbt]
    \centering
    \makebox[\textwidth][c]{
        \begin{tabular}{@{}c@{\hspace{0.5cm}}c@{}}
            \raisebox{-0.5\height}{\includegraphics[height=0.24\textheight]{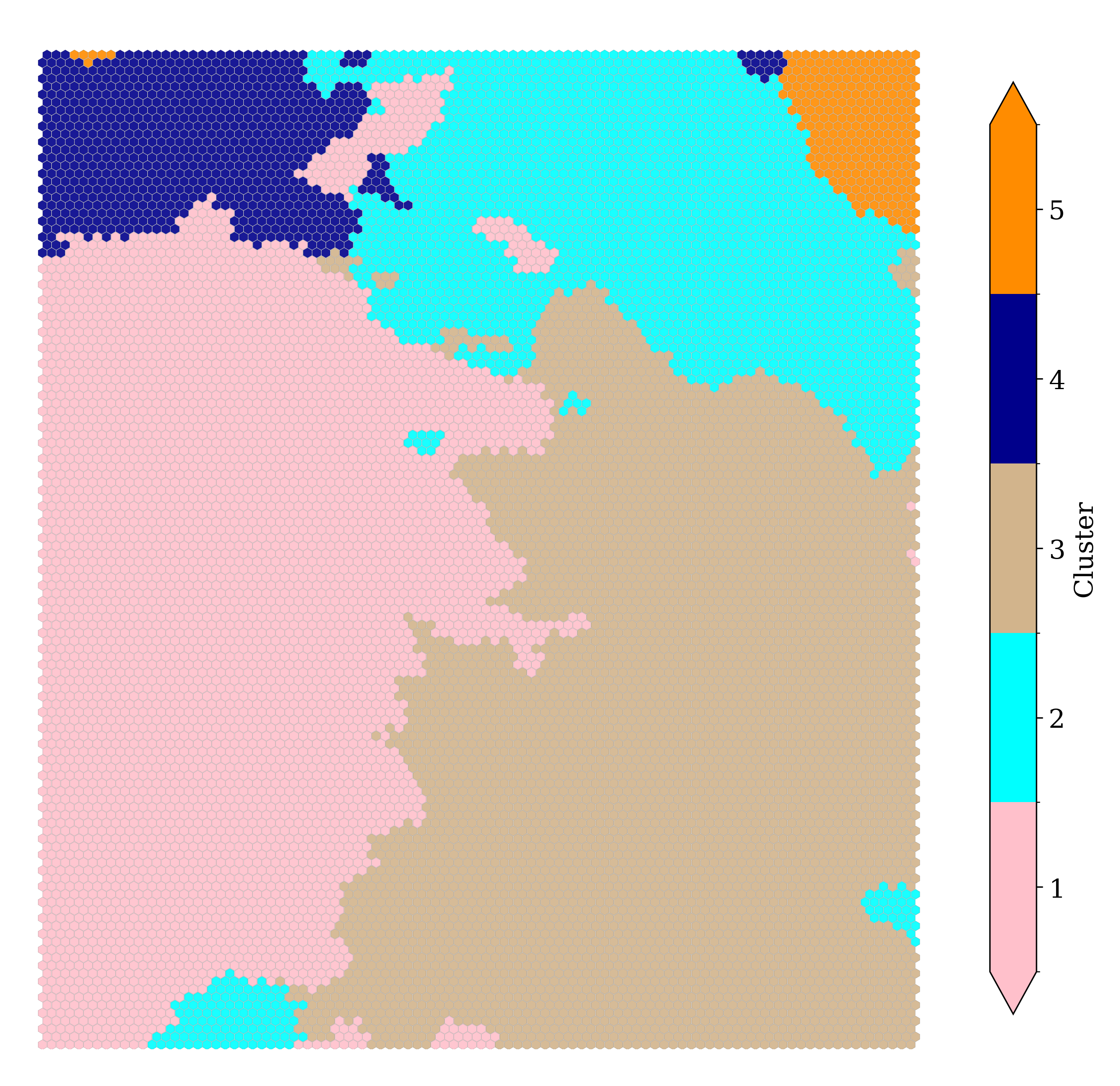}} &
            \begin{tabular}{cc}
                \includegraphics[width=0.23\textwidth]{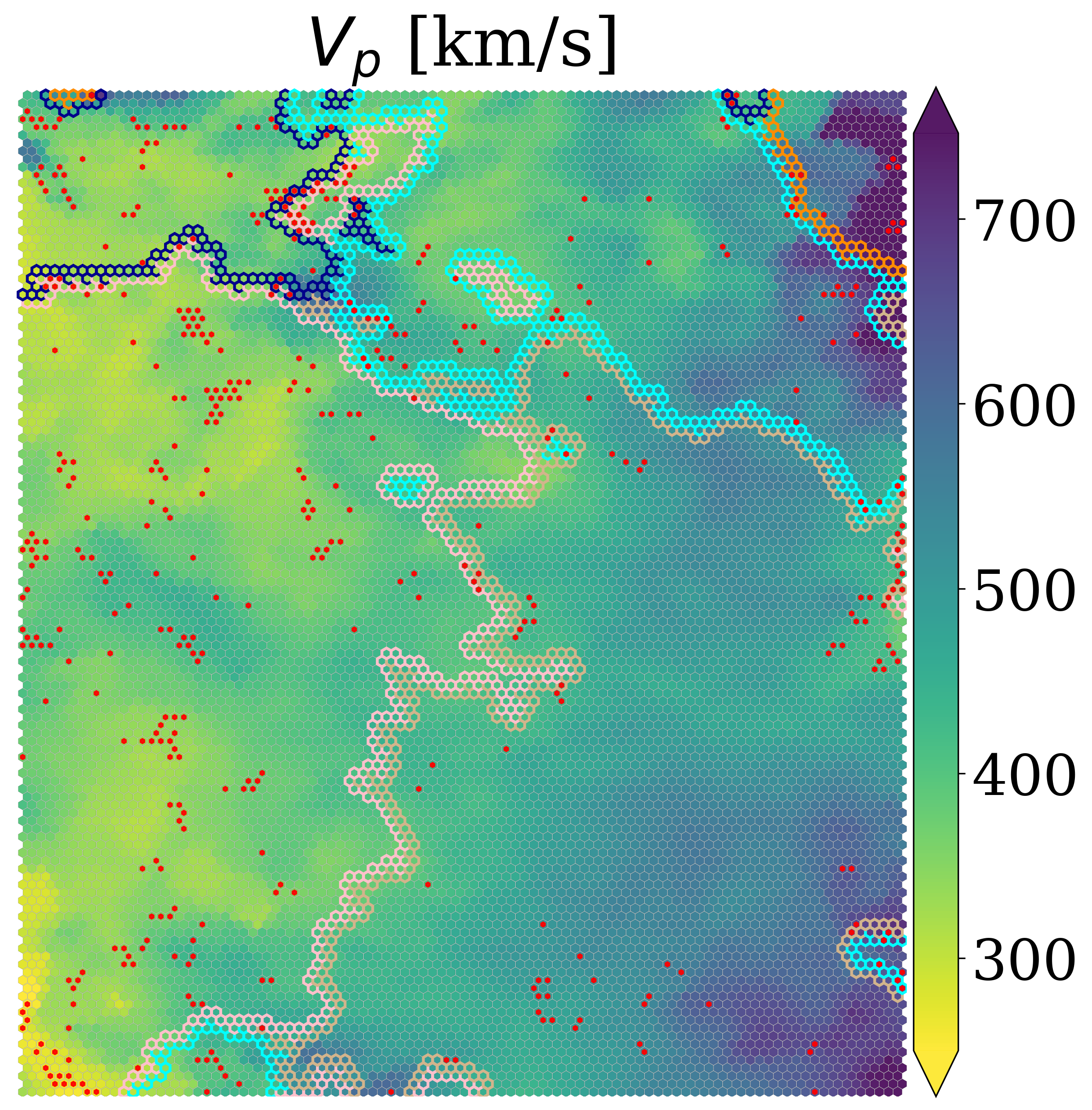} &
                \includegraphics[width=0.22\textwidth]{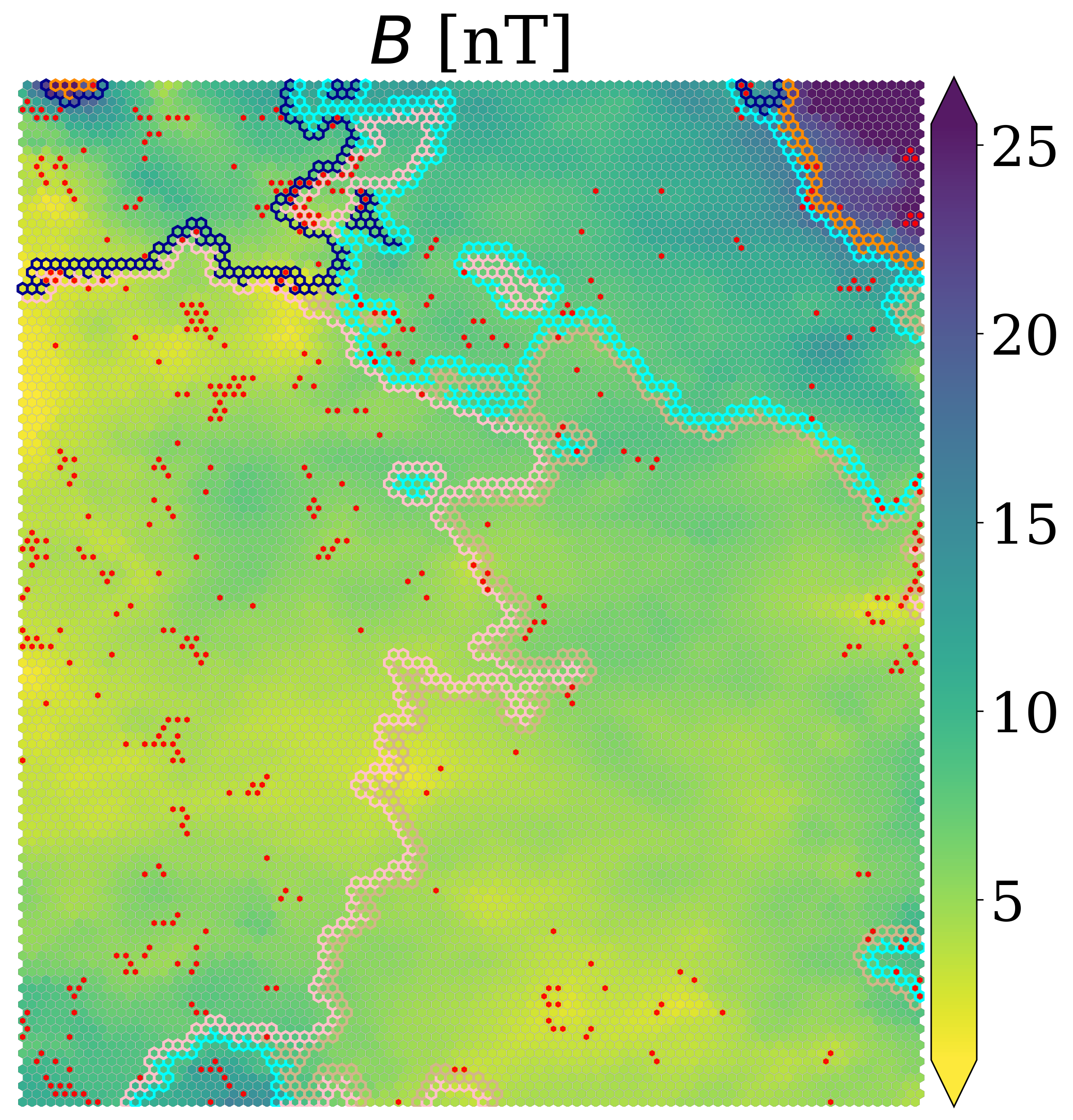} \\
                \includegraphics[width=0.22\textwidth]{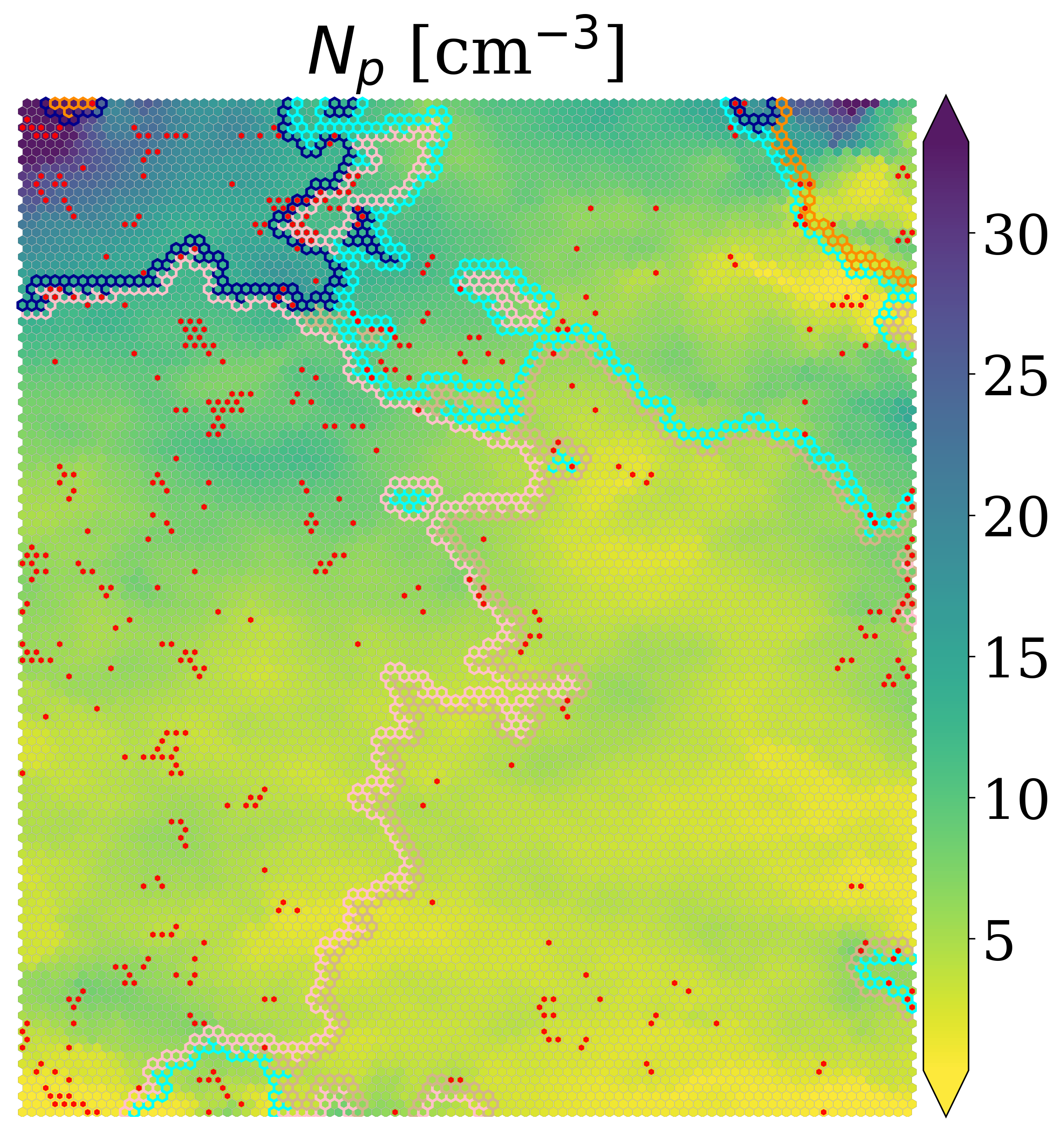} &
                \includegraphics[width=0.23\textwidth]{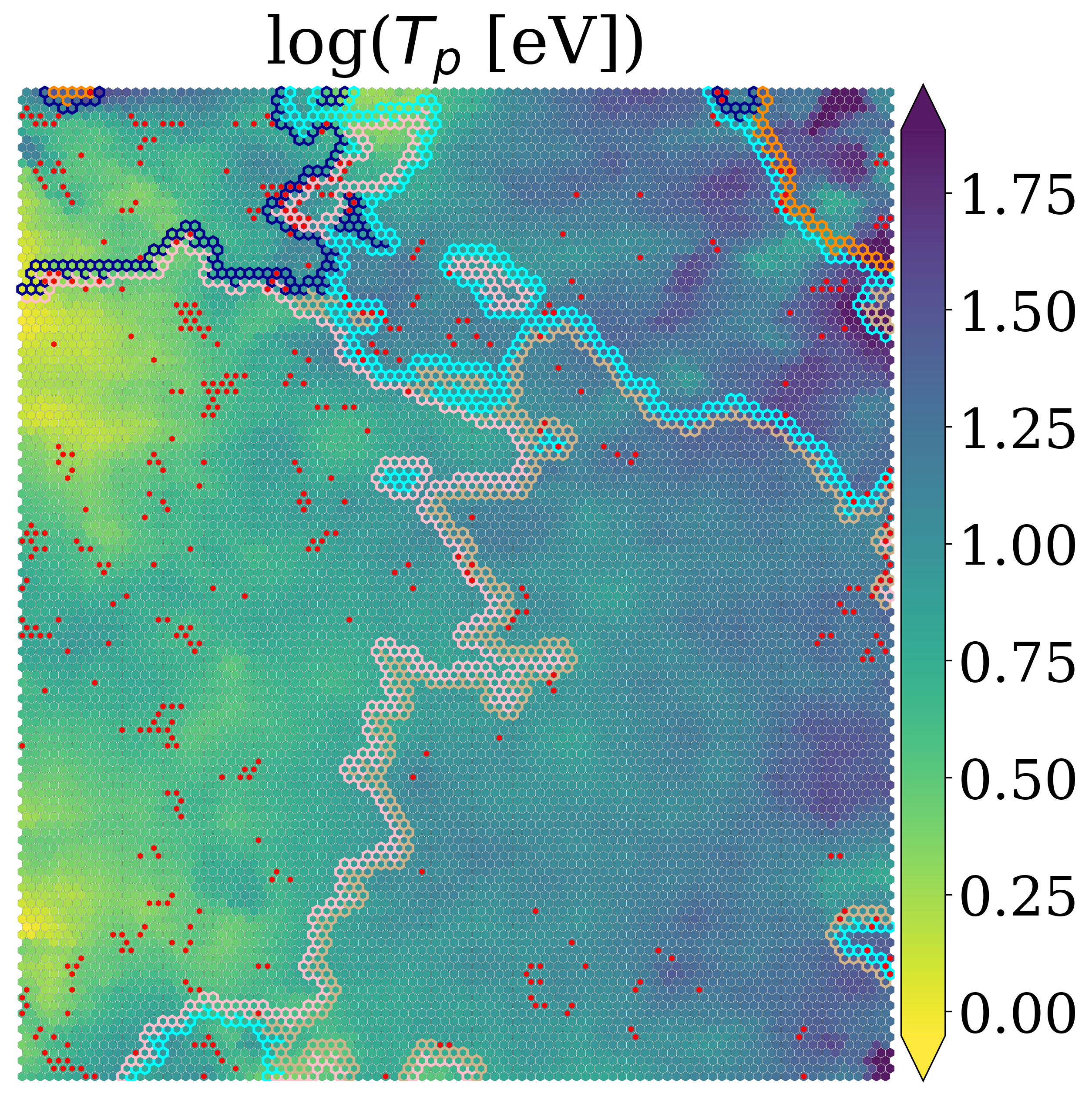} \\
            \end{tabular}
        \end{tabular}
    }
    \caption{(\textit{Left}) Clustered map. (\textit{Right}) Feature maps: proton speed, magnetic field strength, proton density, and proton temperature with overplotted cluster boundaries. In red, we highlight the reconnection neurons.}
   \label{fig:umatrix_kmeans}
\end{figure*}

The optimal number of clusters obtained by applying the Kneedle method to SOM's nodes is \(K\)=5 (see Appendix \ref{appendix:B}). The K-Means clustered map, together with feature maps overplots, is shown in Figure \ref{fig:umatrix_kmeans}. Each node in the map is colored according to the cluster it belongs to.\\
It is possible to observe that the reconnection data are distributed in different clusters, confirming the results inferred by the previous inspection of the feature maps.\\
From the overplots of the clusters' boundaries on the top of the feature maps (Figure \ref{fig:umatrix_kmeans}, on the right) it is possible to extract some properties of the clusters: in Cluster 1 we observe a low proton speed, low temperature, and average density; in Cluster 2 we observe an enhanced proton speed, and a higher temperature and magnetic field strength compared to Cluster 1; in Cluster 3 we observe a high proton speed, high temperature, and low density; in Cluster 4 we observe a high proton density, low velocity, low temperature, and medium to low magnetic field strength; in Cluster 5 we observe a strong magnetic field, high temperature, and high proton speed.
\subsection{Cluster analysis and physical interpretation}
\begin{figure*}[!h]
	\centering
	\includegraphics[width=0.84\linewidth]{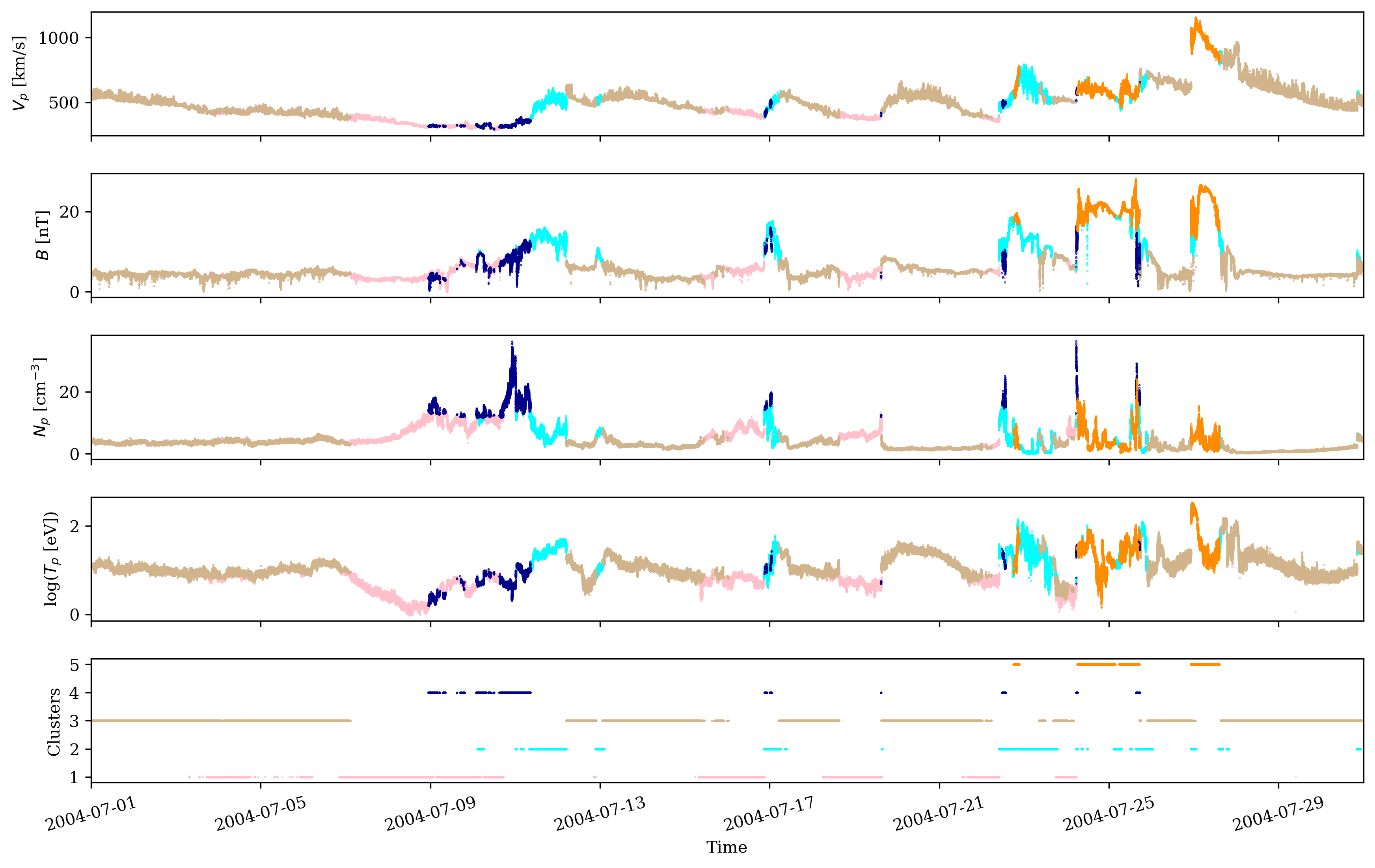}
   \caption{Plot of the clustered time series for July 2004. Each color corresponds to a cluster identified by the SOM+K-Means method: pink is SSW, cyan is CFW, brown is HAW, blue is CSW, and orange is Ejecta.}	
   \label{fig:tskmeans}
\end{figure*}

\begin{figure*}[!h]
	\centering
	\includegraphics[width=0.84\linewidth]{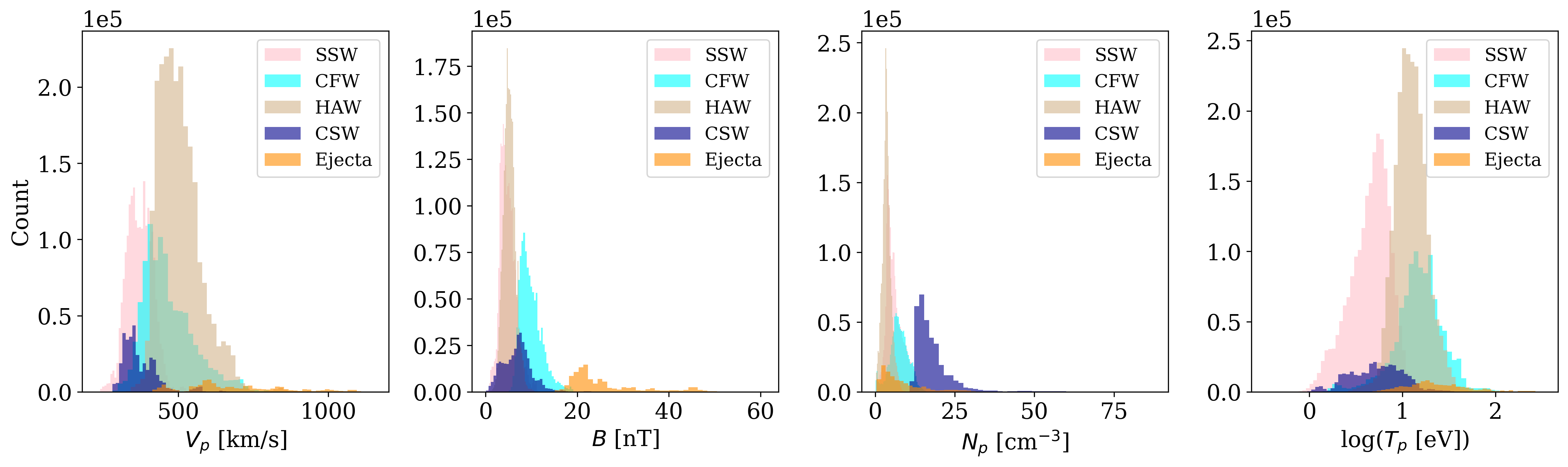}
   \includegraphics[width=0.84\linewidth]{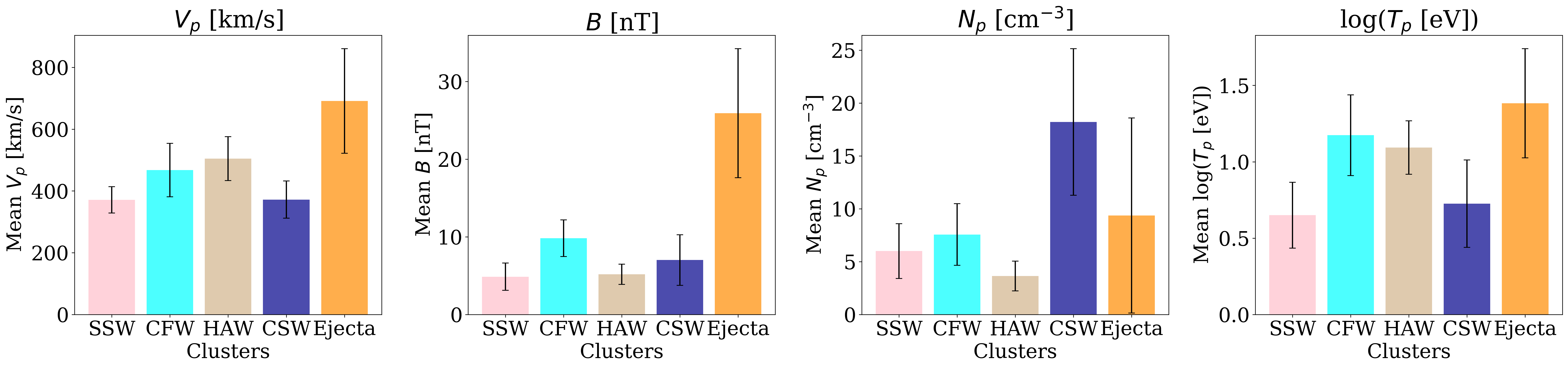}
	\caption{(\textit{Top}) Feature distribution for each cluster. The x axis represents the feature values, while the y axis represents the number of occurrences of that value in the cluster. (\textit{Bottom}) Average values of each feature for each cluster. The error bars represent the standard deviation of the feature values in each cluster.}	
   \label{fig:clustdist}
\end{figure*}

Each node of the map can be considered as a local average of data points with similar properties, meaning that different parts of the time series can be mapped onto each neuron. Therefore, it is possible to associate each solar wind data point with one of the five solar wind clusters.\\
In Figure \ref{fig:tskmeans}, we plot the month of July 2004 as an example of a clustered time series.\\
The blue cluster (Cluster 4) is associated with a part of the time series where the proton density is high, the magnetic field strength is increasing, and the solar wind speed is low. We associate this cluster with a compression of the slow solar wind, as it happens in conjunction with the early phase of transient events, especially SIRs, where slow wind is compressed by an incoming fast flow.  The pink cluster (Cluster 1) is associated with part of the time series where the proton speed is low, the proton density is high, and the temperature is low. We identify it as slow solar wind. The brown cluster (Cluster 3) is associated with part of the time series where the proton temperature is high, the solar wind speed is high, and the density is low: fast solar wind. The cyan cluster (Cluster 2) is associated with part of the time series where the proton temperature presents high values and the proton density is higher with respect to Cluster 1 and 3, along with the magnetic field strength, whereas the velocity presents higher values compared to Cluster 4 and 1. This can be associated with a compressed fast wind, which is typical in SIRs and ICMEs' sheath region. The orange cluster (Cluster 5) is associated with extreme transients in the solar wind where all the features present high values, except for the temperature: we associate this cluster with CME ejecta.\\
In Figure \ref{fig:clustdist}, we plot, at the top, the feature distributions for each cluster, and at the bottom, the average values of each feature for each cluster, which confirms our visual inspection of the time series. Therefore, we decided to assign the following physical interpretation to the clusters: 
\begin{itemize}
   \item Cluster 1 (Pink): \textit{Slow solar wind (SSW)}.
   \item Cluster 2 (Cyan): \textit{Compressed fast wind (CFW)}. 
   \item Cluster 3 (Brown): \textit{Highly Alfvénic wind (HAW)}.
   \item Cluster 4 (Blue): \textit{Compressed slow wind (CSW)}.
   \item Cluster 5 (Orange): \textit{Ejecta}.
\end{itemize}
The reason we decided to assign the name HAW to the brown cluster is explained in more detail in Section \ref{sec:alfvenic}.

\subsection{Reconnection data within the clusters}
\begin{figure*}[!ht]
   \centering
   \begin{minipage}[b]{0.3\linewidth}
      \centering
      \includegraphics[width=\linewidth]{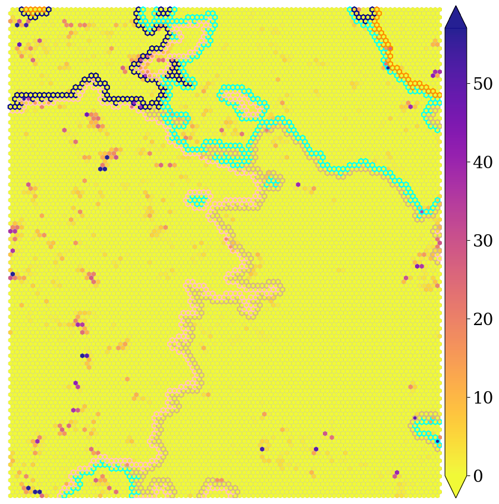}
   \end{minipage}
   \begin{minipage}[b]{0.33\linewidth}
      \centering
      \includegraphics[width=\linewidth]{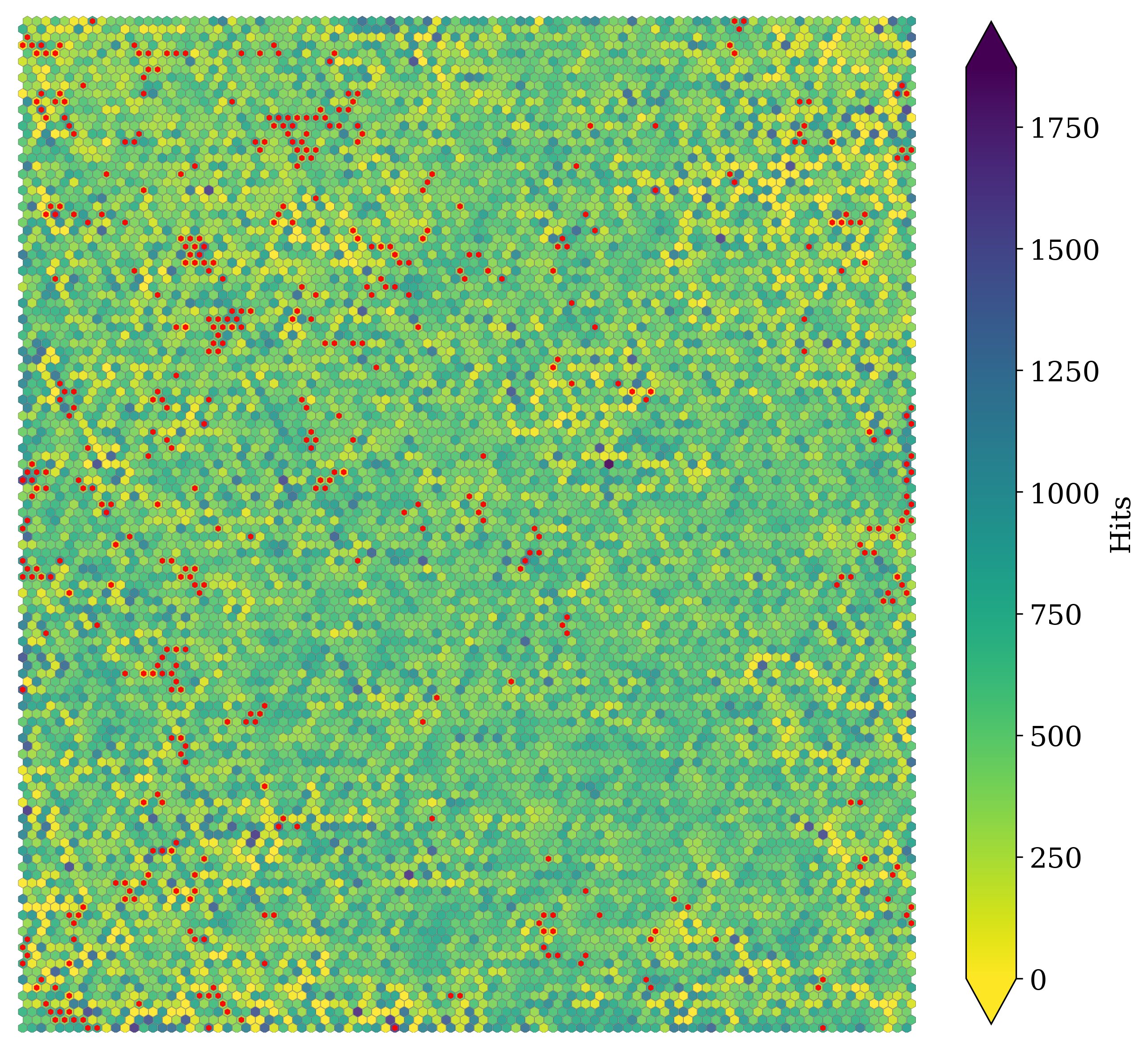}
   \end{minipage}
    \caption{(\textit{Left}) Hitmap of the reconnection data on the SOM nodes with the clusters boundaries overplotted: SSW (pink), HAW (brown), CSW (blue), CFW (cyan), and Ejecta (orange). (\textit{Right}) Hitmap of all data on the SOM nodes.}	
    \label{fig:reconnmap}
\end{figure*}

Now that we have a physical interpretation of the clusters,  we can analyze the distribution of the reconnection data across them. \\
We first performed a visual inspection of the distribution of the reconnection data across the map.\\
In Figure \ref{fig:reconnmap}, on the left, we plot the hitmap of the reconnection data on the SOM's nodes with the clusters boundaries overplotted: for each SOM node, we plot how many reconnection events were detected among the data that have that node as the BMU. On the right, as a comparison, we plot the hitmap for all data, i.e., how many observations are associated with each neuron. It is possible to see that the majority of reconnection events are located in the SSW cluster (pink), followed by the CSW cluster (blue) and the CFW cluster (cyan). The Ejecta cluster (orange) is the one associated with the lowest number of reconnection events. The HAW has the lowest number of activated reconnection neurons, but most of them have a high number of hits. Some reconnection neurons are clustered in a region within this cluster where the velocity is around 450 km/s (Figure \ref{fig:hexvhighlight}, in the yellow box), and the temperature is higher than in the SSW cluster. The hitmap of all data shows that the number of hits is, on average, equally distributed across the map.
After the visual inspection, we quantified the distribution of the reconnection data across the clusters. In Table \ref{tab:reconnection_clusters} we show the number of reconnection events in each cluster and the relative reconnection content, which is the percentage of reconnection events with respect to the total number of data points in the cluster.\\
The table confirms that the majority of the reconnection events (more than 50\%) are located in the SSW cluster, followed by the HAW cluster (around 17\%), the CSW cluster (around 15\%), and the CFW cluster (around 14\%). The Ejecta cluster (around 3\%) has the lowest percentage of reconnection data. It is noticeable that each cluster has a low relative reconnection content, less than 1\%, due to the fact that the total number of reconnection points (7756) is low compared to the size of the initial dataset. However, the clusters with the highest relative reconnection content are the CSW and the Ejecta, followed by SSW, CFW, and HAW.
\begin{table*}[!ht]
   \centering
   \caption{Distribution of reconnection events within each cluster obtained with SOM + K-Means.}
   \label{tab:reconnection_clusters}
   \renewcommand{\arraystretch}{1.2}
   \begin{tabular}{|c|c|c|c|}
      \hline
      \rowcolor{gray!25}
      Cluster & Reconnection within cluster (\%) & Relative reconnection content (\%) \\
      \hline
      \rowcolor{gray!5}
      \cellcolor{mypink}{\color{white} SSW (Slow Solar Wind)} & 51.19 & 0.18 \\
      \hline
      \rowcolor{gray!5}
      \cellcolor{mycyan}{\color{white} CFW (Compressed Fast Wind)} & 13.68 & 0.10 \\
      \hline
      \rowcolor{gray!5}
      \cellcolor{mytan}{\color{white} HAW (Highly Alfvénic Wind)} & 17.36 & 0.06\\
      \hline
      \rowcolor{gray!5}
      \cellcolor{myblue}{\color{white} CSW (Compressed Slow Wind)} & 15.03 & 0.33 \\
      \hline
      \rowcolor{gray!5}
      \cellcolor{myorange}{\color{white} Ejecta} & 2.73 & 0.19 \\
      \hline
   \end{tabular}
   \tablefoot{
   The “Reconnection within cluster” column displays the percentage of reconnection data points in each cluster, while “Relative reconnection content” represents the percentage of reconnection data points in each cluster relative to the total number of data points in the cluster.
    }
\end{table*}
\subsection{The highly Alfvénic wind cluster}
\label{sec:alfvenic}
\begin{figure}[!hbt]
   \centering
   \includegraphics[width=0.92\linewidth]{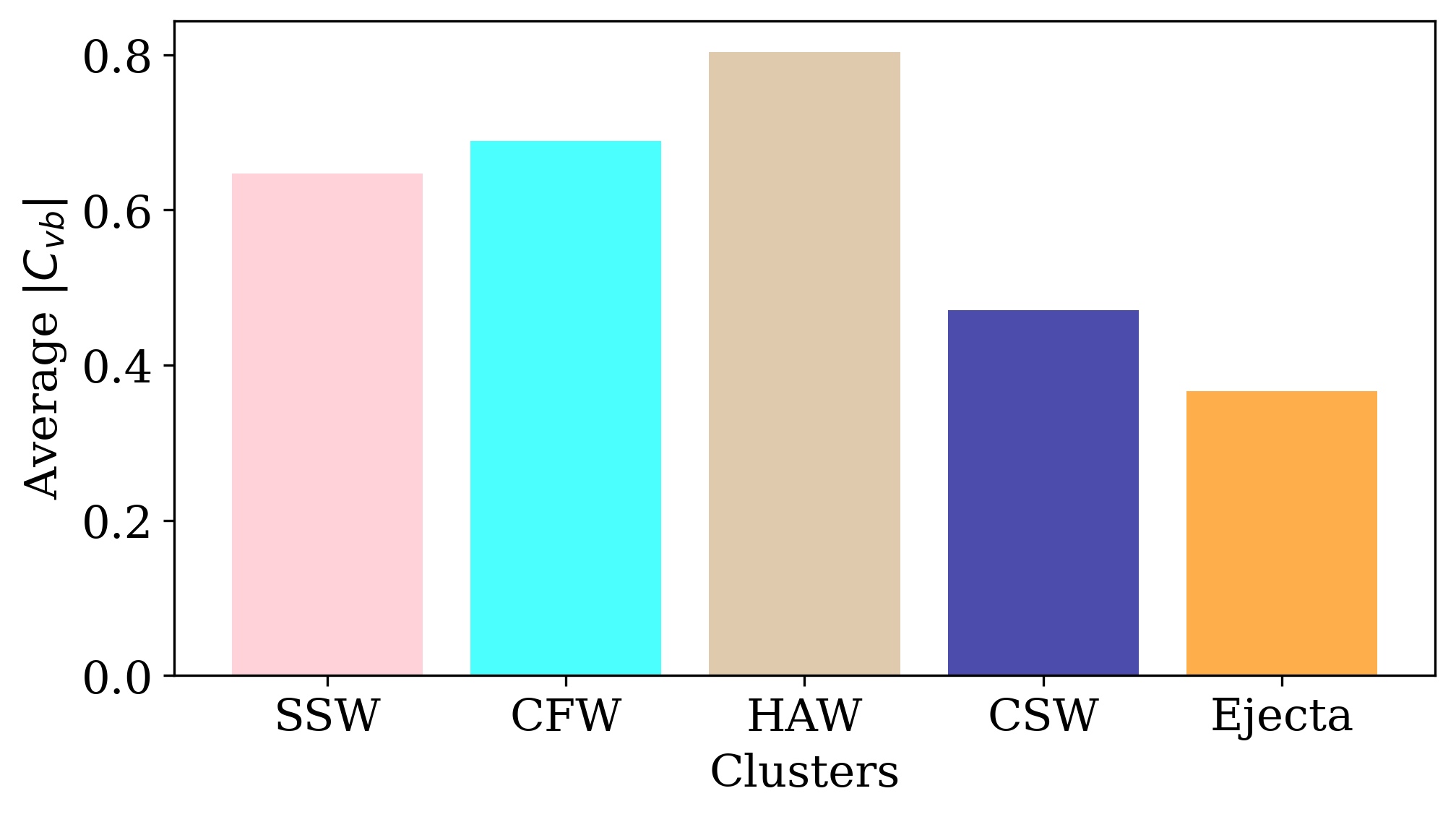}
   \caption{Average absolute value of the correlation coefficient, $C_{vb}$, for each cluster. The HAW cluster (brown) has a high average absolute value of the correlation coefficient, indicating that the fluctuations of the velocity and magnetic field are strongly correlated (highly Alfvénic wind).}
   \label{fig:Alfvénicity}
\end{figure}

The HAW cluster (brown) has proton speed ranging from around 450 km/s to 700 km/s, which is a range that includes low to high solar wind velocities. To further investigate this cluster, we decided to look at its Alfvénicity.\\
Alfvénicity can be quantified by the correlation coefficient, $C_{vb}$, between the fluctuations of the velocity and magnetic field's components computed using a 30 min running window \citep{damicis_first_2021}, as in:
\begin{equation}
    C_{vb} = \frac{1}{3} \sum_i C_{vb,i} \hspace{0.2cm} \text{with} \hspace{0.2cm}
   C_{vb,i} = \frac{\sum_j (V_{i,j} - \overline{V_{i}})(B_{i,j} - \overline{B_{i}})}{\sqrt{\sum_j (V_{i,j} - \overline{V_i})^2 (B_{i,j} - \overline{B_i})^2}},
\end{equation}
where $V_{i,j}$ and $B_{i,j}$ are the $j$-th samples of the velocity and magnetic field $i$-th component ($i=x,y,z$), and $\overline{V_i}$ and $\overline{B_i}$ are their mean values over a 30-minute time interval.\\
A value of $C_{vb}$ close to 1 (-1) indicates that the fluctuations of the velocity and magnetic field are strongly positively (negatively) correlated, while a value close to 0 indicates that they are not correlated. The Alfvénic wind has a high value for the absolute value of $C_{vb}$.
We computed the average absolute value of the correlation coefficient, $C_{vb}$, for each cluster, as is shown in Figure \ref{fig:Alfvénicity}. The HAW cluster has a high average absolute value of the correlation coefficient (0.8), indicating that the fluctuations of the velocity and magnetic field are strongly correlated. This suggests that the plasma in this cluster is highly Alfvénic. We see that the CFW cluster is the second cluster with the highest average absolute value of $C_{vb}$, followed by the SSW cluster, the CSW cluster, and the Ejecta cluster. The Ejecta cluster has the lowest average absolute value of $C_{vb}$, indicating that the fluctuations of the velocity and magnetic field are not correlated here.
\subsection{Comparison with \citet{Xu}}
\label{sec:comparison}
\begin{figure}[!hbt]
   \centering
   \includegraphics[width=0.96\linewidth]{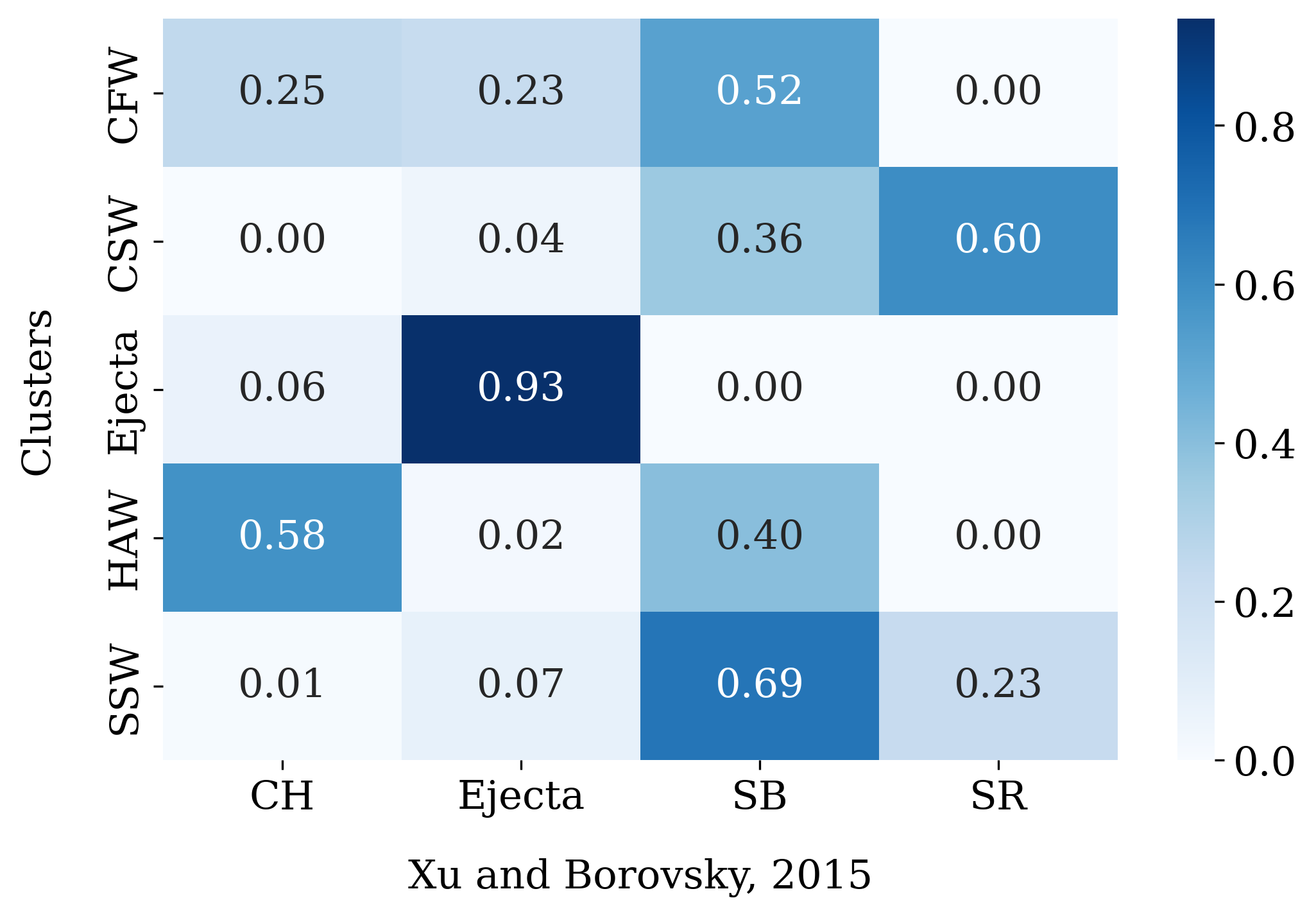}
   \caption{Comparison between the solar wind classification from \citet{Xu} (on the x axis) and our clustering results (on the y axis). Each row shows the percentages of \citep{Xu} classes within the clusters identified by SOM + K-Means.}
   \label{fig:comparison}
\end{figure}

\begin{figure*}[!hbt]
         \centering
         \includegraphics[width=0.84\linewidth]{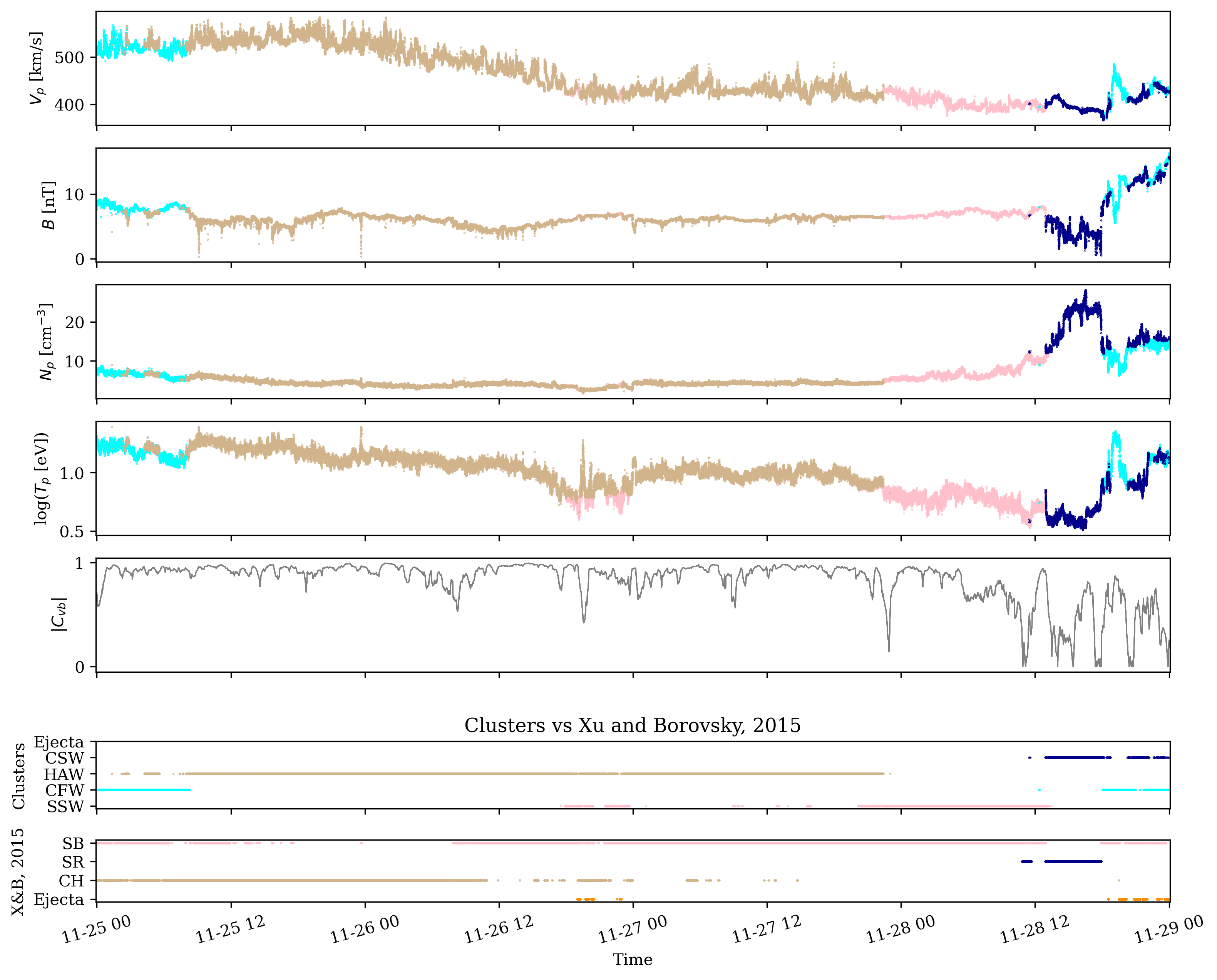}
         \caption{(\textit{Top}) Comparison between our clustering and the Xu and Borovsky classification for a fast flow during November 2004. The Alfvénicity parameter, $|C_{vg}|$, is plotted in black. Our HAW cluster (brown) includes data points classified in \cite{Xu} as both CH and SB, but all are characterized by high Alfvénicity.}
         \label{fig:alfvennov}
\end{figure*}

\begin{figure*}[!hbt]
         \centering
         \includegraphics[width=0.84\linewidth]{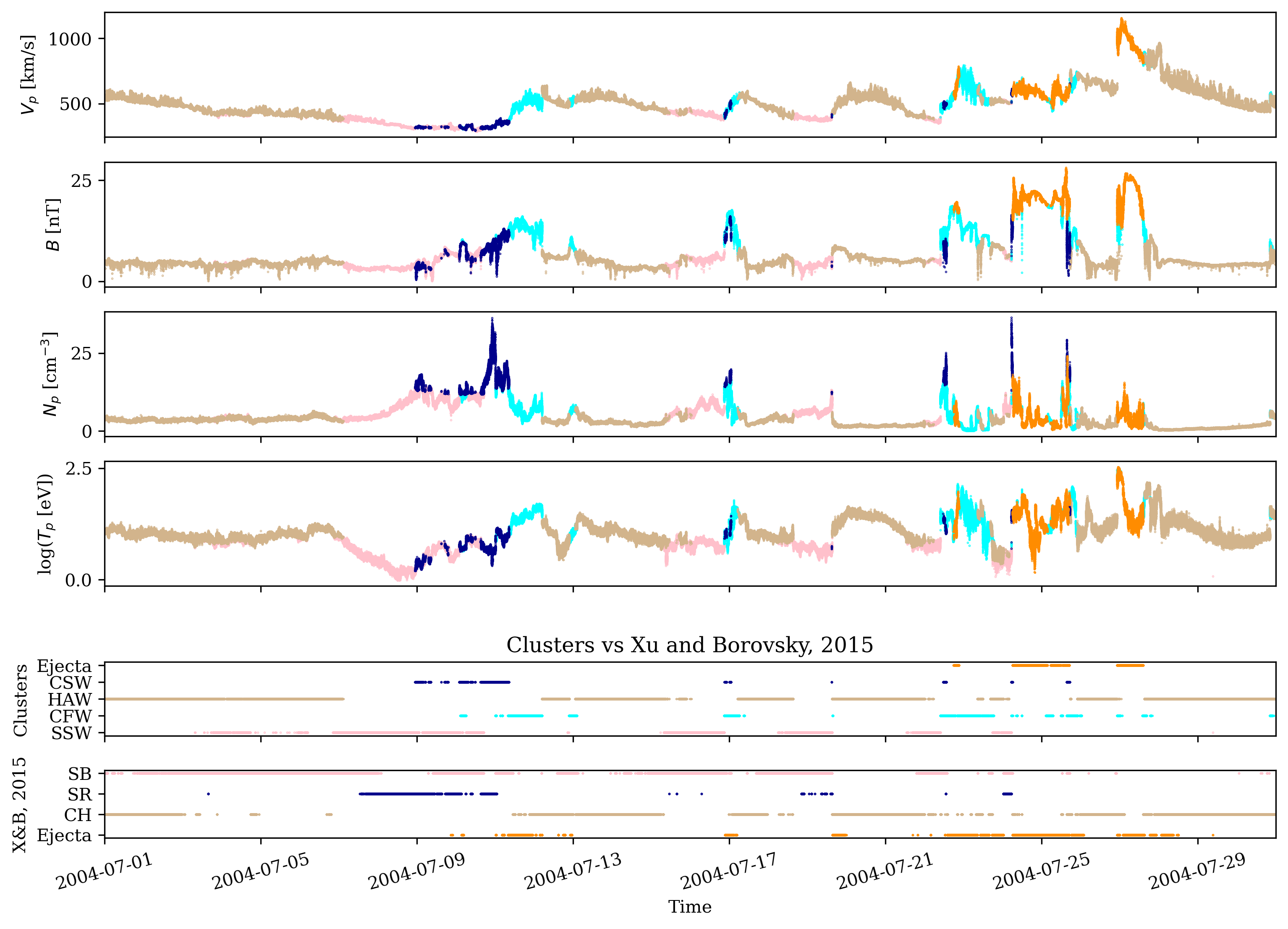}
         \caption{(\textit{Top}) Clustered time series for July 2004. (\textit{Bottom}) Comparison between the \citet{Xu} classification and our clustering results. It is noticeable that the Ejecta class in the \citet{Xu} classification is also partially superimposed here on our compressed fast wind, which is associated with a SIR.}
         \label{fig:julyxuvssom}
\end{figure*}

In \citet{Xu}, the authors provided a solar wind classification based on a different set of parameters. The classification is based on the following parameters: the Alfvén speed, \(V_A\), the proton specific entropy, \(S_p\), and the temperature ratio, \(T_{exp}/T_p\), defined as follows:
\begin{equation}
   V_A = \frac{B}{\sqrt{4\pi m_p N_p}}, \quad
   S_p = \frac{T_p}{N_p^{2/3}}, \quad
   T_{exp} = \left(\frac{V_p}{283}\right)^{3.113}
,\end{equation}
where \(B\) is the magnetic field strength, \(m_p\) is the proton mass, \(N_p\) is the proton density, and \(T_p\) is the proton temperature.\\
The classification is aimed at identifying different solar wind types, based on their origin, using the criteria in Table \ref{tab:xbtab}.
\begin{table}[h!]
\centering
\caption{\citet{Xu} classification criteria.}
\begin{tabular}{|p{0.15\columnwidth}|p{0.76\columnwidth}|}
\hline
\textbf{Category} & \multicolumn{1}{c|}{\textbf{Condition}} \\
\hline
Ejecta & 
$\log_{10} V_A > 0.277 \log_{10} S_p + 0.055 \log_{10} \left( \frac{T_{exp}}{T_p} \right) + 1.83$ \\
\hline
Coronal Hole (CH) & 
$\log_{10}(S_p) > -0.525 \log_{10} \left( \frac{T_{exp}}{T_p} \right) - 0.676 \log_{10}(V_A) + 1.74$ \newline
and not classified as Ejecta \\
\hline
Sector Reversal (SR) & 
$\log_{10}(S_p) < -0.658 \log_{10}(V_A) - 0.125 \log_{10} \left( \frac{T_{exp}}{T_p} \right) + 1.04$ \newline
and not classified as Ejecta \\
\hline
Streamer Belt (SB) & Remaining data points not satisfying any of the above conditions \\
\hline
\end{tabular}
\label{tab:xbtab}
\end{table}
Ejecta incorporates magnetic clouds and ICMEs, Coronal Hole (CH) is solar wind originating from coronal holes, i.e., fast wind, Streamer Belt (SB) is solar wind originating from the streamer belts, usually associated with slow solar wind, and Sector Reversal (SR) is wind originating from the sector reversal region, corresponding to the helmet streamer regions (see Figure 1 in \cite{Xu}). 
We applied the same classification to our dataset and compared the results with our clustering results. The results are shown in Figure \ref{fig:comparison}. On the y axis, we have the clusters identified from our method, while on the x axis, we have the classification from \citet{Xu}. We then observed how the data from each of ``our" clusters would be classified according to \cite{Xu} (the sum of each row is 1). We observed that our SSW cluster mainly maps to streamer belt plasma and partially to sector reversal plasma in \cite{Xu}, as was expected given the possible region of origin of slow solar wind plasmas.  Two ejecta clusters correspond well. Our CSW cluster mostly maps to sector reversal plasma and partially to streamer belt plasma. The HAW cluster is mainly composed of coronal hole plasma, as was expected, followed by streamer belt plasma. This may be for two different reasons. The first one, more based on statistics, is that the cadence in our dataset is very high (3s), while the thresholds obtained in \citet{Xu} are based on a lower-cadence dataset (1-hour averages), and this can also explain why we see mismatches in the previous clusters; because the HAW cluster includes not only the ``usual" fast, highly Alfvénic solar wind plasma, but also some highly Alfvénic solar wind plasma with velocities around 450 km/s (see Figure \ref{fig:alfvennov}). The CFW cluster's composition is the one that captures more attention: according to the \citet{Xu} clustering, it is composed of coronal hole plasma, streamer belt plasma, and ejecta plasma. This is partially explained by the conclusions of \cite{Xu}, in which the authors mention that they noticed a bimodal distribution of their ejecta plasma: one with magnetic-cloud-like patterns of carbon-oxygen ratios, and the other with coronal-hole-like and streamer-belt-like patterns of carbon-oxygen ratios. From Figure \ref{fig:julyxuvssom}, it is possible to notice that the identified ejecta plasma by \citet{Xu} partially superimposes on our compressed fast wind plasma, which is a wind typically occurring right after a compressed slow wind in an SIR. A typical SIR is characterized by an initial compression of the slow wind, followed by a region of compressed fast wind, which comes before the high-speed stream, as is illustrated in Figure 13 in \citep{bd1971}.\\
We remark that different SIR events may have different signatures, which may result in having only compressed fast wind or compressed solar wind plasma signatures. If we look at the two SIRs in Figure \ref{fig:tskmeans}, respectively, one SIR on 9-13 July and the other on 18-23 July, we note that the first SIR is characterized by a sequence of both compressed slow and fast plasma, whereas the second one is characterized by only compressed slow plasma. The reason why we see these signatures may need further investigation. It may depend on how the speed of the incoming flow, how compressed the slow wind is, and how the spacecraft trajectory is directed with respect to the incoming SIR. One should not forget that SIRs and ICMEs are 3D structures, and the in situ signatures can differ depending on the trajectory of the spacecraft with respect to the incoming structure.
\newline
To further investigate the differences between our clustering results and those from \citet{Xu}, we analyzed the distribution of reconnection events in the clusters identified following the \citet{Xu} classification. The results are shown in Table \ref{tab:reconn_xu}.\\
Most of the reconnection data are located in the SB cluster (50.86\%), followed by the SR cluster (26.65\%) and the Ejecta cluster (14.82\%). The CH cluster has the lowest percentage of reconnection data (7.66\%). This is mostly compatible with our results in Table \ref{tab:reconnection_clusters}. We see that the reconnection event distribution percentage in the SSW cluster is compatible with that in the SB wind in Table \ref{tab:reconn_xu}. If we sum the reconnection content of our CFW and Ejecta cluster, we get a comparable percentage to the Ejecta plasma in \citet{Xu}, whereas there are some mismatches with the SR plasma and the CH plasma. The reason for this is partially explained by the result in Figure \ref{fig:comparison}, where we see that the HAW cluster contains a 40 \% amount of SB plasma, and the CSW cluster has 60 \% of SR plasma. The relative reconnection content mostly matches what we obtained with our method, with the exception of the HAW cluster, for the reasons we mentioned above. \\ Overall, the results that we obtain are compatible with the ones in \citet{Xu}. The mismatches are due to the fact that we aimed for a solar wind clustering highlighting the presence of transients (SIR, CMEs, ...) superimposed on the background flow, while \citet{Xu} aimed for a solar wind classification based on the solar wind origin. The two approaches are complementary and could potentially be used together to achieve a better understanding of the occurrence of magnetic reconnection in the solar wind.
\begin{table*}[!hbt]
   \centering
   \caption{Distribution of reconnection events within the solar wind classification from \citet{Xu}.}
   \label{tab:reconn_xu}
   \renewcommand{\arraystretch}{1.2}
   \begin{tabular}{|c|c|c|c|}
      \hline
      \rowcolor{gray!25}
      Solar Wind Origin & Reconnection within cluster (\%) & Relative reconnection content (\%) \\
      \hline
      \rowcolor{gray!5}
      \cellcolor{mypink}{\color{white} SB (Streamer Belt)} & 50.86 & 0.13 \\
      \hline
      \rowcolor{gray!5}
      \cellcolor{myblue}{\color{white} SR (Sector Reversal)} & 26.65 & 0.29\\
      \hline
     \rowcolor{gray!5}
      \cellcolor{mytan}{\color{white} CH (Coronal Hole)} & 7.66 & 0.04 \\
      \hline
      \rowcolor{gray!5}
      \cellcolor{myorange}{\color{white} Ejecta} & 14.82 & 0.20 \\
      \hline
   \end{tabular}
\end{table*}
\section{Conclusions}
\label{sec:conclusions}
In this work, we have applied unsupervised learning techniques, specifically SOM \citep{som3} and K-Means clustering, to the analysis of solar wind data at 1 AU. Our goal was to identify patterns in the data and understand the occurrence of magnetic reconnection events in the solar wind and its transients. \citet{HEIDRICHMEISNER2018397}, \citet{bloch2020data}, and \citet{amaya2020visualizing} have used a similar approach to analyze solar wind data at 1 AU from the ACE mission \citep{ACE1997}, with 12 min cadence data.\\ Here we clustered solar wind data from the Wind spacecraft at a 3-second cadence. We used Wind data because the magnetic reconnection catalog in \citet{eriksson2022} is based on Wind observations \citep{ogilvie1997}. We adopted the 3-second cadence because it has been shown that the median value for the reconnection exhaust duration is around 12 seconds; the 3-second cadence allows us to better define the reconnection exhausts and to have more data associated with them.\\ 
We started by collecting magnetic field data from the MFI instrument \citep{1995} and proton density, proton temperature, and solar wind speed data from the 3DP instrument \citep{3dp} on the Wind spacecraft. We focused on six months of data from the year 2004, which was covered by the reconnection exhausts catalog from \citet{eriksson2022}. After preprocessing the data by removing low-quality data points and applying robust scaling \citep{Rousseeuw}, we trained a SOM to transform the time series data into a visual map.\\
The trained SOM allowed us to visualize the high-dimensional data in a lower-dimensional space, making it easier to understand the underlying patterns and structure of the solar wind data. We then applied the K-Means clustering algorithm to the trained SOM to cluster the SOM nodes and identify distinct clusters within the solar wind data. The optimal number of clusters (5) was determined using the Kneedle algorithm \citep{satopaa}. This was further confirmed with the Calinski-Harabasz scores \citep{chi} and the Silhouette analysis \citep{ROUSSEEUW198753} (see Appendix \ref{appendix:B}).\\
We identified clusters associated with slow solar wind, compressed slow solar wind, highly Alfvénic wind, compressed fast wind, and ejecta. Our analysis revealed that the reconnection observations are unevenly distributed across the different clusters. The majority of reconnection events were found in the slow solar wind cluster. The slow solar wind originates mostly from streamer belt regions, which are associated with closed magnetic field lines and are therefore more likely to generate complex current sheets that can potentially reconnect. We found that 17\% of the reconnection data is associated with a highly Alfvénic wind: this is less expected, but since this cluster includes solar wind with velocities around 450 km/s, this matching is likely since this is a transition wind between fast and slow wind, and therefore it is possible to have potentially reconnecting current sheets in these time intervals. The rest of the reconnection data are associated with compressed slow and fast solar wind, with a small fraction of data associated with ejecta. \\
These findings suggest that magnetic reconnection events in the solar wind occur under different solar wind conditions, and the majority are associated with the slow solar wind and time intervals where the solar wind is compressed, which is likely to happen during an encounter with an SIR or an ICME. Whether compressed fast wind or compressed slow wind is present or not during such an SIR or ICME depends on many factors, such as the type of event or from which direction the event is approaching the spacecraft or vice versa. In Figure \ref{fig:icmesjuly} we show an example of how the different clusters are associated with four ICMEs from the \citet{nieves2} catalog, showing that each event has different characteristics. This is also mentioned in \citet{HEIDRICHMEISNER2018397}, where ICMEs were excluded. 
The comparison with the solar wind classification from \citet{Xu} showed that our clustering results and reconnection event distribution results are compatible with their classification if we consider that the aim of this work is not to classify the solar wind based on its origin, but rather to have clusters associated with the solar wind conditions in the near-Earth environment. We found that our slow solar wind cluster is mostly composed of streamer belt plasma, while the compressed slow solar wind cluster is mostly composed of sector reversal plasma, and both of them have very low percentages associated with coronal hole plasma or ejecta plasma. We found that the compressed fast wind cluster is a mixture of streamer belt plasma, coronal hole plasma, and ejecta plasma, and this may explain the observed bimodal distribution of ejecta plasma mentioned in the conclusions of \citet{Xu}, where it was found that the ejecta are composed of two populations: one with magnetic-cloud-like patterns of carbon-oxygen ratios, and the other with coronal-hole-like and streamer-belt-like patterns of carbon-oxygen ratios. The highly Alfvénic wind cluster is mostly composed of coronal hole plasma, but it also includes a significant amount of streamer belt plasma, which is likely because this cluster includes a solar wind with velocities around 450 km/s, which is a range that includes intermediate solar wind, as was previously mentioned. Another possible explanation is that \citep{Xu} used lower-cadence data (1-hour averages) compared to our dataset (3 seconds), and this can lead to different statistical thresholds that can lead to different classification results.\\
Overall, the combination of SOM and K-Means clustering proved to be an effective tool for visualizing and analyzing solar wind observations, allowing us to identify physically explainable clusters and to contextualize the occurrence of magnetic reconnection events in the solar wind. Since the SOM algorithm tends to diverge, the SOM results might need to be verified independently. With the comparison with \citet{Xu} and the study in Appendix \ref{appendix:B}, we have verified that the SOM provided useful and trustworthy insights. 
While we focus on a rather limited time interval in a specific phase of the solar cycle, the clusters and the reconnection event distribution in the clusters that we obtain are compatible with the \citet{Xu} classification, which is based on a wider temporal range of solar wind data.
This work poses a basis for a future extended study specifically aimed at characterizing solar wind transients. This could be achieved by reducing the cadence of the dataset to include more information from different solar cycles.
\newpage
\begin{acknowledgements}
   The author acknowledges support from the European Research Council within the ERC Advanced Grant TerraVirtualE (ERC-2022-ADG, Project reference: 101095310). M.E.I. and S.K. acknowledge support from the Deutsche Forschungsgemeinschaft (German Research Foundation) SFB1491. The authors acknowledge the usage of the Wind data downloaded from the CDAWeb database. The authors acknowledge the usage of Python libraries such as SunPy, Sklearn, Pandas, Numpy, Matplotlib, and Kneed for data analysis and clustering, and the CUDA-SOM tool for the SOM training. The main author acknowledges the usage of GitHub Copilot as an accessory tool to help improving the code and write code to speed up the manuscript's writing. 
   Finally, the main author would like to thank his former supervisor, Prof. Giovanni Lapenta, who sadly passed away in May 2024, for his guidance and support throughout this research, Dr. Luca Franci for the discussions on the Alfvénic solar wind, Dr. George Miloshevich for their general discussions, and the referees who significantly contributed to improving the quality of this manuscript.
\end{acknowledgements}

\bibliographystyle{aa}
\bibliography{bibliography}

\begin{appendix}
   \section{SOM Parameters}
      \label{appendix:A}
      Here we discuss the parameters choice for the SOM training.\\ \citet{som3} suggests the following formula for the number of neurons $K$ in a map:
      \begin{equation}
         K = 5 \times \sqrt{N}
      \end{equation}
      where \(N\) is the number of data points in the dataset.\\
      To set the dimension of the map, we have to consider the number of neurons \(K\) and the number of rows \(x\) and columns \(y\) of the map, such that $x \cdot y \approx K$.\\ To do so, we decide that the ratio between the two dimensions should be the same as the ratio of the two principal components \citep{pearson} of the dataset:
      \begin{equation}
         \frac{x}{y} = \frac{\text{PC1}}{\text{PC2}}  
      \end{equation}
      where PC1 and PC2 are the first two principal components' eigenvalues.\\
      In this work, the decaying function for both neighborhood radius $\sigma$ and the learning rate $\eta$ is the following:
      \begin{equation}
         f(\tau) = f_0 \exp\left(-\frac{\tau}{\sqrt{T_{max}}}\right)
      \end{equation}
      where $f_0$ is the initial value of the parameter, $\tau$ is the iteration step and $T_{max}$ is the maximum number of iterations.\\
      $T_{max}$ is chosen so that toward the end of the training, the neighborhood radius approaches 1, meaning that at the end of the training, the updates are more local and the overall map structure is less impacted by them. Given the initial value of the learning rate $\eta_0$ and the critical value $\eta_{crit}$, the upper value number of maximum iterations $T_{max}$ is given by:
      \begin{equation}
         T_{max} \leq log\left(\frac{\sigma_{crit}}{\sigma_0}\right)^2 = log\left(\frac{\eta_{crit}}{\eta_0}\right)^2
      \end{equation}
      We decide to set the initial learning rate to $\eta_0 = 0.01 \cdot \sigma_0$ because we do not want to have large updates mostly affected by large learning rates during the training (see Eq. \ref{eq:wupd}) and a maximum number of epochs $T_{max}$=9.

    \section{Comparison with K-Means and Convergence}
    \label{appendix:B}
\begin{figure}[!hbt]
            \centering
            \includegraphics[width=\linewidth]{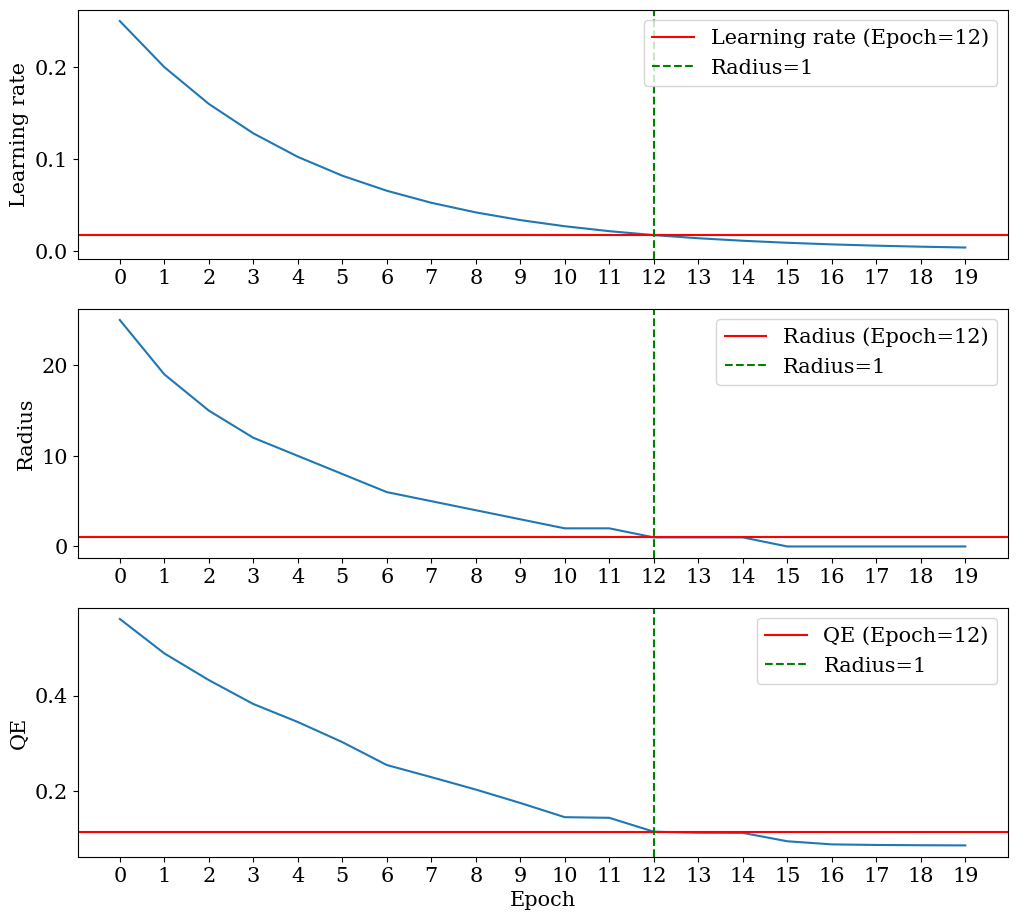}
            \caption{Trend of the learning rate, the radius, and the quantization error during the SOM's training. In red is highlighted the baseline for each quantity when the radius is equal to 1. Green-dashed is the value of the epochs such that the radius first approaches 1.}
            \label{fig:epochs}
        \end{figure}

\begin{figure*}[!t]
            \centering
            \includegraphics[width=0.32\linewidth]{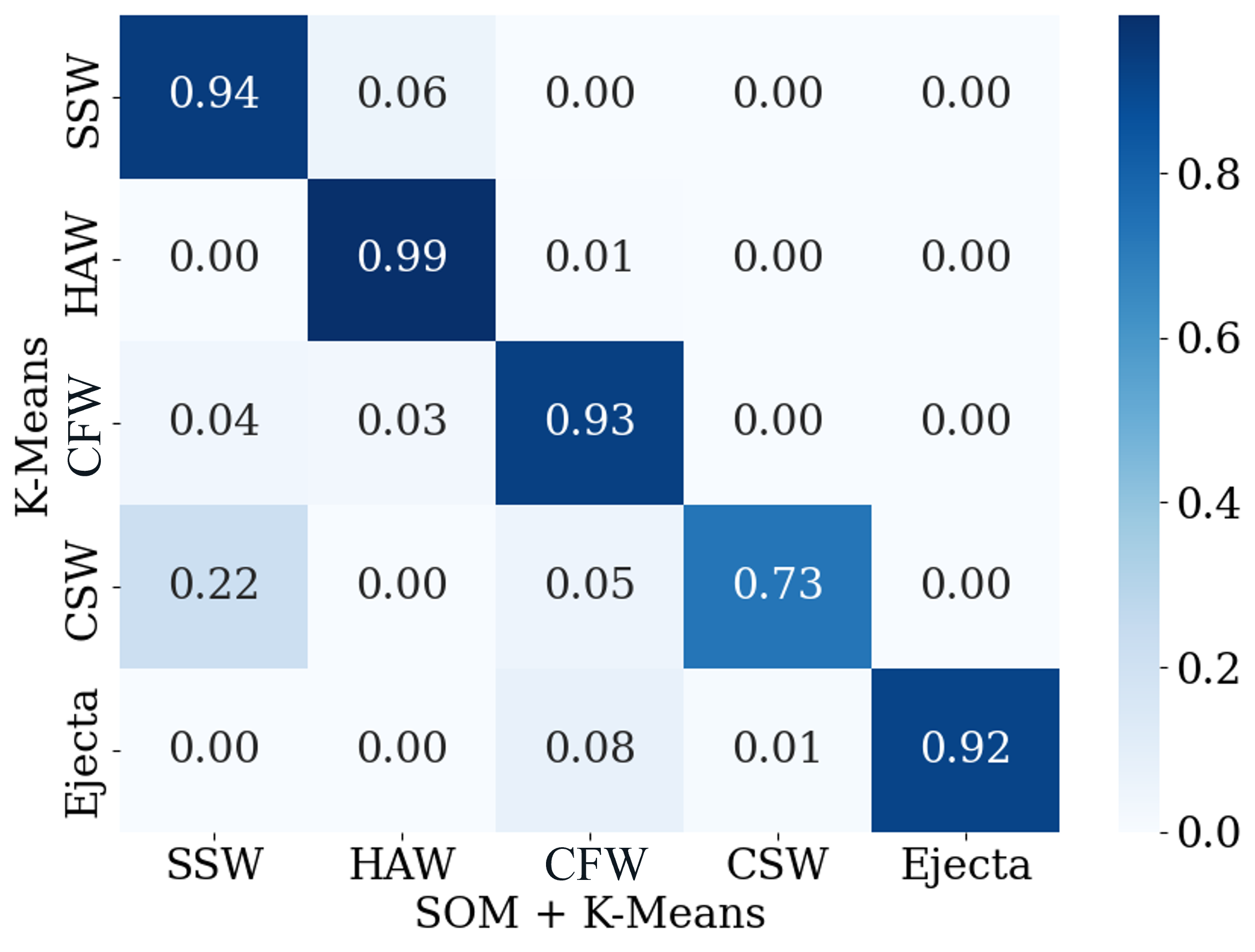}
            \includegraphics[width=0.32\linewidth]{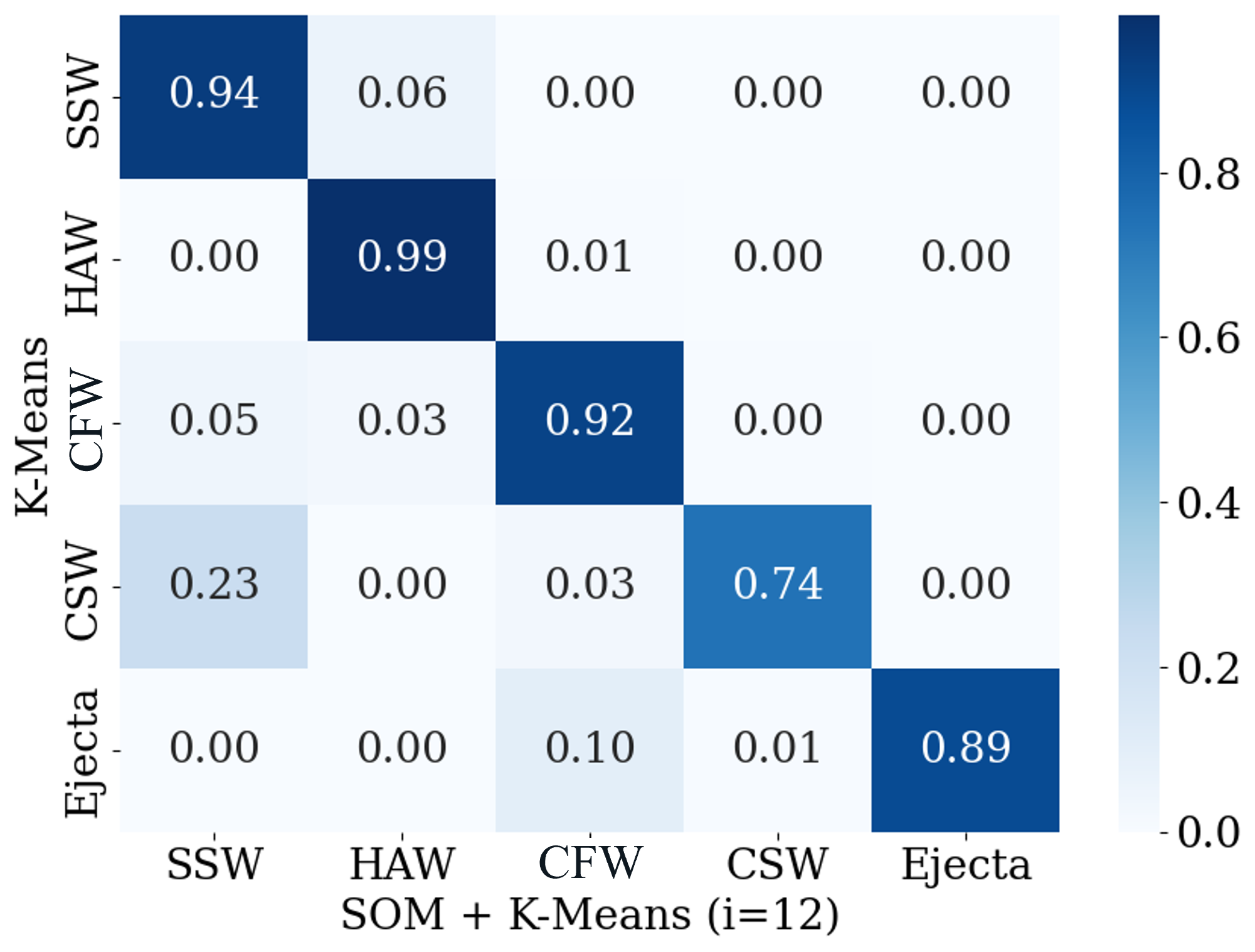}
            \includegraphics[width=0.32\linewidth]{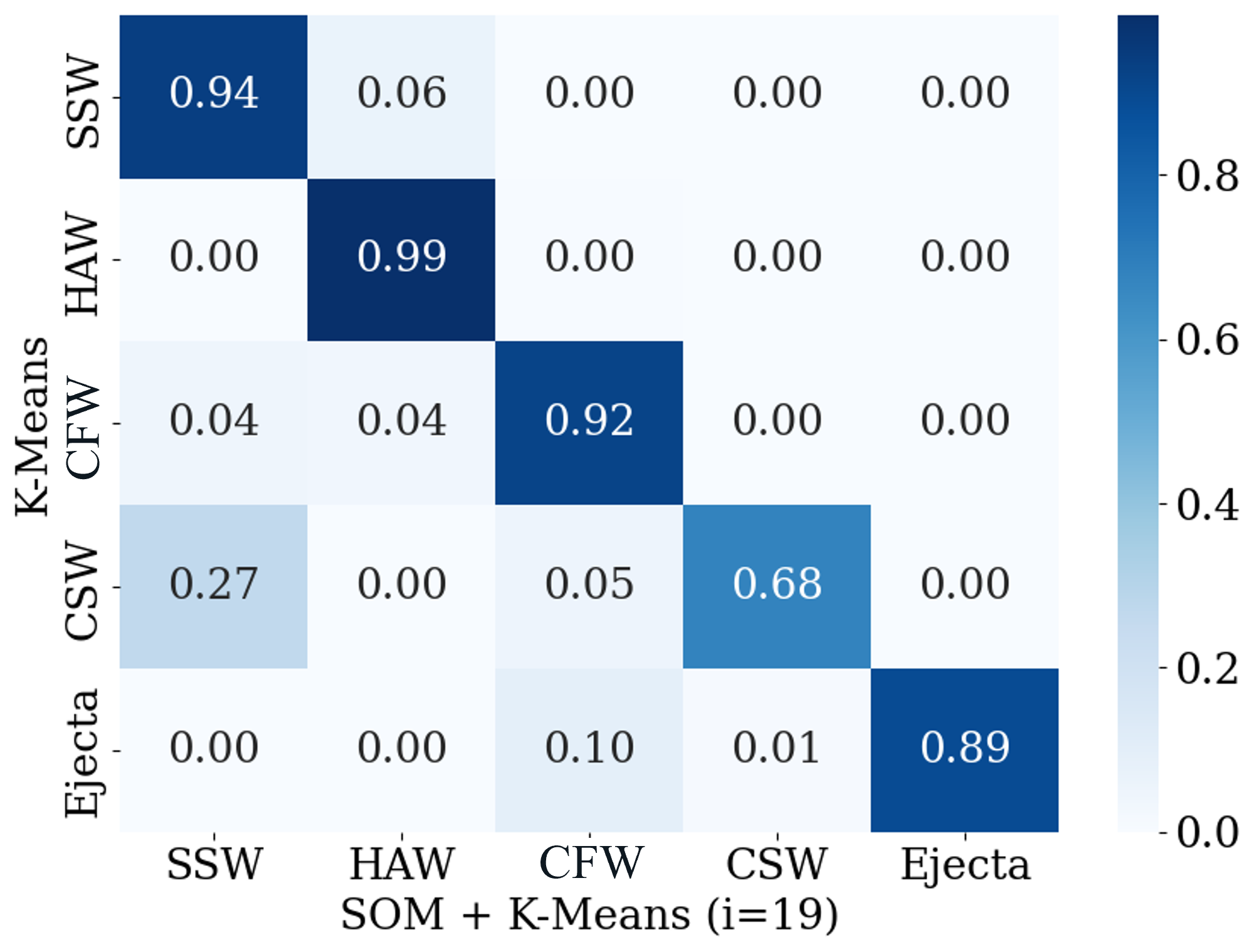}
            \caption{Comparison between K-Means and SOM+K-Means at different stages of the training. From left to right: the SOM that has been used in the main results, long training at epochs equal to 12, and long training at epochs equal to 19.}
            \label{fig:vskmeans}
        \end{figure*}

      To better justify our choice of $T_{max}$, we show how classification results are impacted by a different choice of $T_{max}$.\\
      In the body of the manuscript, we use $T_{max}=9$. Here, we train SOM for 20 epochs. In Figure \ref{fig:epochs}, it is shown how the hyperparameters and the quantization error (QE) of the SOM is changing according to the epochs, where in red is highlighted the epochs for which, with $T_{max}$=20, the radius goes to 1. It is noticeable that after the radius goes to 1, the improvements on the QE are marginal.\\
      After this, we apply K-Means on the data without applying SOM, and compare the results of applying K-Means alone and the clustering results using the SOM that has been used in the main results, and the longer-trained SOM at the stage when the radius is equal to 1 ($i$=12, where $i$ is the epoch) and at the end of the training ($i$=19).
      Figure \ref{fig:vskmeans} shows the comparison between K-Means and SOM + K-Means with the total number of epochs used for the main results ($T_{max}$ = 9 epochs) and at the different training stages in the longer training for 12 epochs, 19 epochs. It is noticeable that the comparison gets worse once we look at the last stages of SOM, and that the results with the SOM stopped at epochs equals to 12 and the ones we used in the main results are compatible. We notice that the main differences between applying K-Means alone and the combination of SOM with K-Means are mostly related to the CSW cluster. The CSW cluster is one of the smallest clusters in the dataset (see Figure \ref{fig:clustdist}), and we see that in the case of SOM + K-Means, we classify it as SSW. We think that this is a minor issue, and it may be related to the fact that with high cadence, slightly different thresholds may result in slightly different clustering. We see that the biggest clusters, SSW and HAW, are the ones that vary the least. 
      \subsection{Training with a different number of neurons}
\begin{figure*}[!t]
            \centering
            \includegraphics[width=0.33\linewidth]{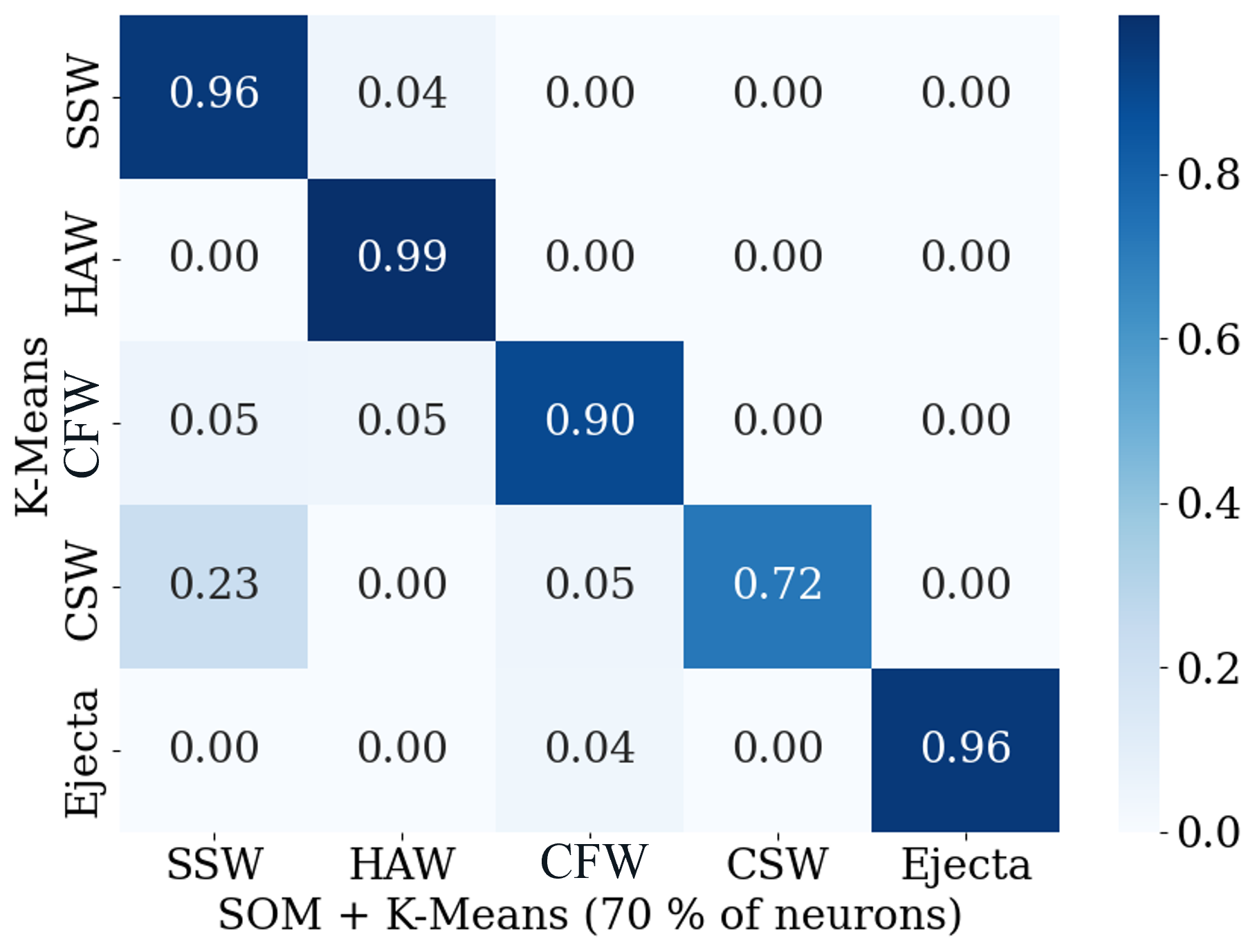}
            \hspace{0.5cm}
            \includegraphics[width=0.33\linewidth]{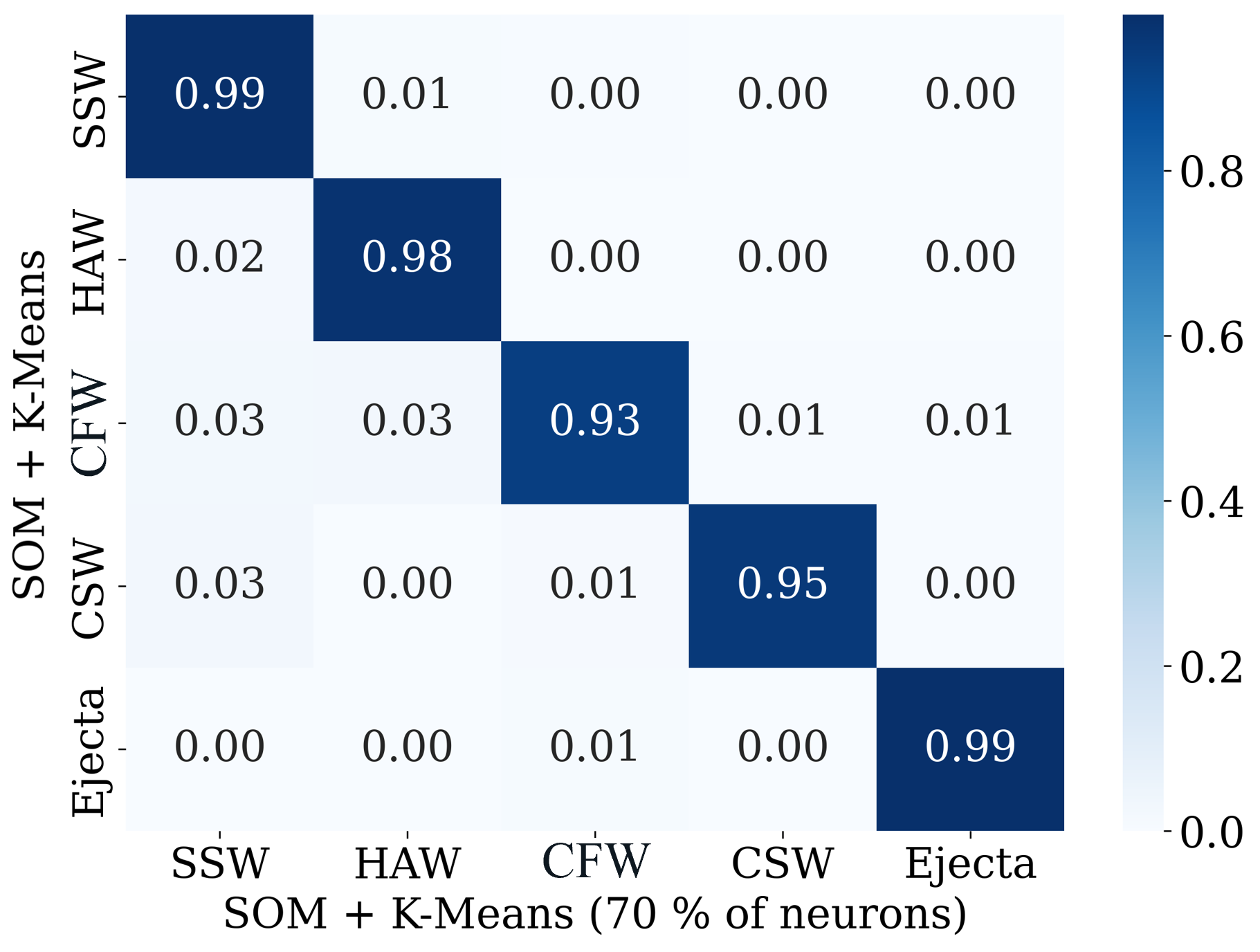}
            \caption{(\textit{Left}) Comparison between K-Means and SOM + K-Means trained with fewer neurons. (\textit{Right}) Comparison between SOM + K-Means with a SOM with fewer neurons and the SOM that has been used in the main results.}
            \label{fig:vskmeans_dn}
        \end{figure*}

      To further justify our choice of parameters we decided to perform a training of the SOM with a different number of neurons. We decide to use 70\% of the neurons used in the main results, and we rescale each hyperparameter accordingly, with $x_{dim}$=105, $y_{dim}$=80, $\sigma_0$=21, $\eta_0$=0.21, $T_{max}$=7.\\
      In Figure \ref{fig:vskmeans_dn}, on the left, we show the comparison with only applying K-Means on the data and the SOM trained with fewer neurons, and on the right, we show the comparison between the SOM used in the main results and the one trained with fewer neurons. It is noticeable that we have similar results with the ones we obtained by comparing K-Means and SOM + K-Means with the SOM used in the main results. This is confirmed by the comparison between the two SOMs in Figure \ref{fig:vskmeans_dn}, on the right, where we have above 93\% matching between clusters.
    
      \section{Finding the optimal number of clusters}
      \label{appendix:C}
         \subsection{Kneedle Method}
\begin{figure}[!hbt]
            \centering
            \includegraphics[width=\linewidth]{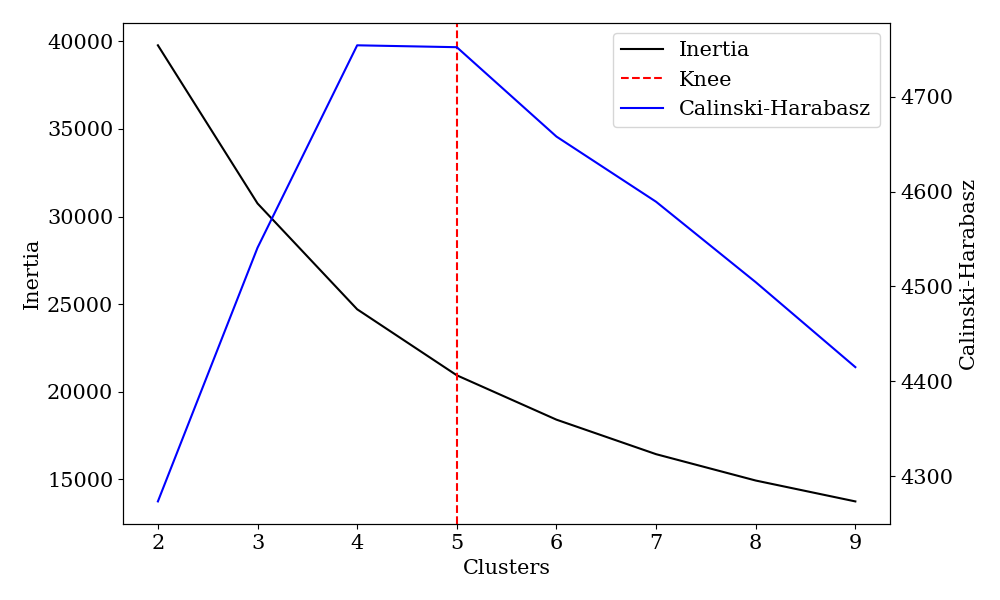}
            \caption{Plot of the inter-cluster variance (in black) and the Calinski-Harabasz (CH) scores (in blue). The knee (red dashed line) identified with the kneedle method corresponds to the maximum CH score.}
            \label{fig:kneedle}
         \end{figure}

         The Kneedle method is a technique to find the optimal number of clusters in a dataset. It is based on the computation of the second derivative of the intra-cluster variance as a function of the number of clusters. The intra-cluster variance, or inertia, is defined as the sum of squared distances between each point and the centroid of the cluster to which it has been assigned:
         \begin{equation}
             \text{Inertia} = \sum_{i=1}^{k} \sum_{x \in C_i} \| x - \mu_i \|^2
         \end{equation}
         The optimal number of clusters is the one that maximizes the second derivative of the intra-cluster variance.\\ Here we show the results of the Kneedle method applied to the SOM's nodes. In this work, we have used the \textit{kneed} \citep{arvai2020kneed} python package to compute the knee of the inertia.\\
         A metric that can be used to evaluate clustering methods is the Calinski-Harabasz index \citep{chi}. The Calinski-Harabasz index (CH) is defined as the ratio of the sum of inter-cluster dispersion and the sum of intra-cluster dispersion. The index is computed as
         \begin{equation}
            \text{CH}(K) = \frac{\text{Tr}(B_K)}{\text{Tr}(W_K)} \times \frac{N-K}{K-1}
         \end{equation}
         where \(N\) is the number of data points, \(K\) is the number of clusters, \(\text{Tr}(B_K)\) is the trace of the intra-cluster dispersion matrix and \(\text{Tr}(W_K)\) is the trace of the inter-cluster dispersion matrix, defined as follows:
         \begin{equation}
            B_K = \sum_{k=1}^K n_k (\Vec{\mu}_k-\Vec{\mu})(\Vec{\mu}_k-\Vec{\mu})^T
         \end{equation}
         where \(n_k\) is the number of data points in the k-th cluster, \(\Vec{\mu}_k\) is the centroid of the k-th cluster, and \(\Vec{\mu}\) is the centroid of the dataset. The intra-cluster dispersion matrix is defined as
         \begin{equation}
            W_K = \sum_{k=1}^K \sum_{\Vec{x} \in C_k} (\Vec{x}-\Vec{\mu}_k)(\Vec{x}-\Vec{\mu}_k)^T
         \end{equation}
         where \(C_k\) is the k-th cluster.\\
         Usually, the optimal number of clusters is the one that maximizes the CH index.\\
         The plot of the intra-cluster variance (inertia) as a function of the number of clusters, together with the CH scores in Figure \ref{fig:kneedle}, shows the match between the knee identified by the Kneedle method and the maximum of the CH index.\\

         \begin{figure*}[!t]
            \centering
            \includegraphics[width=\linewidth]{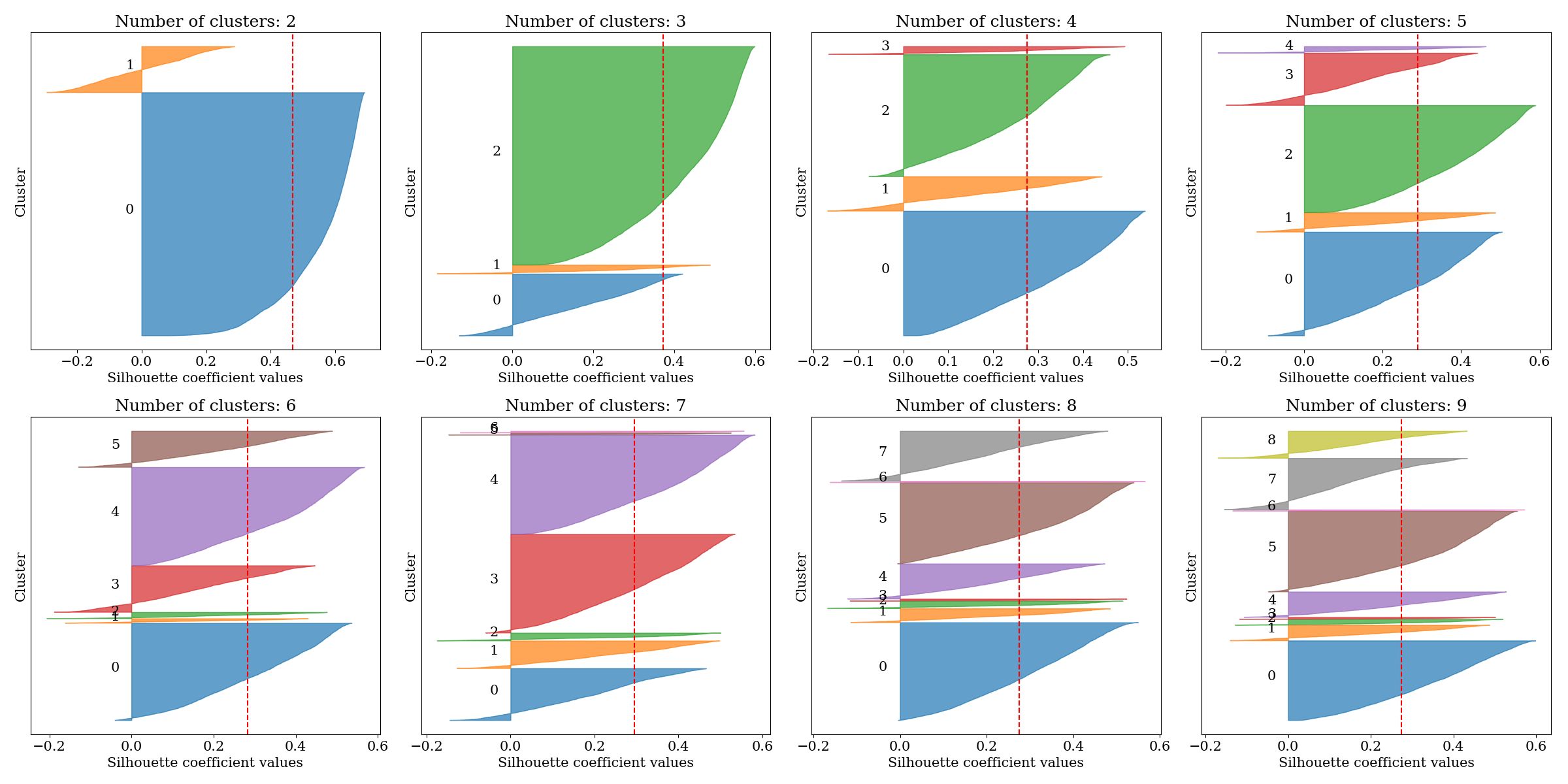}
            \caption{Results of the Silhouette analysis of the K-Means clustering of the trained SOM nodes, with different numbers of clusters. Each plot shows the Silhouette scores for each SOM node in each cluster and the average Silhouette score (red dashed). The optimal number of clusters is 5 because all the clusters' scores are above the average Silhouette score which is greater than the rest of the average scores with different numbers of clusters, excluding the case of \(K\)=2, where one cluster (orange) has lower scores than the average and \(K\)=3, where two clusters (cyan and orange) are slightly above the average.}
            \label{fig:silhouette}
         \end{figure*}

         \subsection{Silhouette Analysis}
         Another technique to find the optimal number of clusters is the silhouette analysis. The silhouette score is a measure of how similar an object is to the other data points in its cluster compared to those in other clusters. The silhouette score ranges from -1 to 1, where a value close to 1 indicates that the object is well-matched to its cluster, negative values indicate that an object may have been assigned to the wrong cluster, and values close to 0 indicate that the object is lying close to the decision boundaries between catalog clusters. The silhouette score for a single object \(i\) is computed as
         \begin{equation}
            s(i) = \frac{b(i)-a(i)}{\max(a(i),b(i))}
         \end{equation}
         where \(a(i)\) is the average distance between \(i\) and all other data points in the same cluster, and \(b(i)\) is the minimum average distance between \(i\) and all other clusters.\\
         The silhouette analysis consists of measuring the silhouette score for each data point and then computing the average silhouette score for each number of clusters. The optimal number of clusters is a compromise between the maximization of the average silhouette score and how much each cluster's scores are above the average silhouette score.\\
         In Figure \ref{fig:silhouette}, we show the silhouette analysis for the clustering of the SOM's nodes: the silhouette score for each SOM node and the average silhouette scores are shown. The conditions expressed above are satisfied for \(K=5\), thus confirming the results obtained with the Kneedle method.
      \clearpage
      \section{Additional Figures}
      \label{appendix:D}
      \begin{figure}[!hbt]
	      \centering
	      \includegraphics[width=\linewidth]{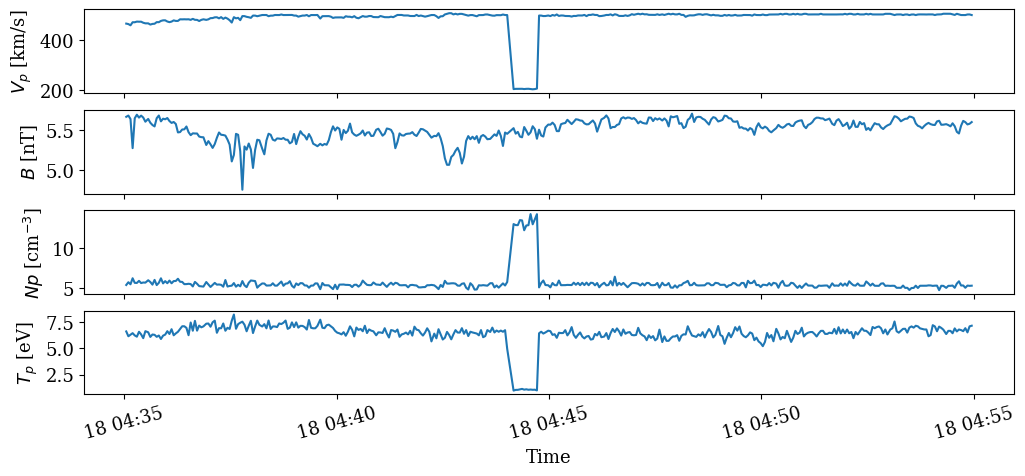}
	      \caption{Example of bad quality data window for September 2004. It is noticeable a sudden drop of the velocity at 200 km/s. The other features of the 3DP instrument are affected by the same drop or jumps at the same time.}
	      \label{fig:badata}
      \end{figure}
      \begin{figure}[!hbt]
	      \centering
	      \includegraphics[width=0.9\linewidth]{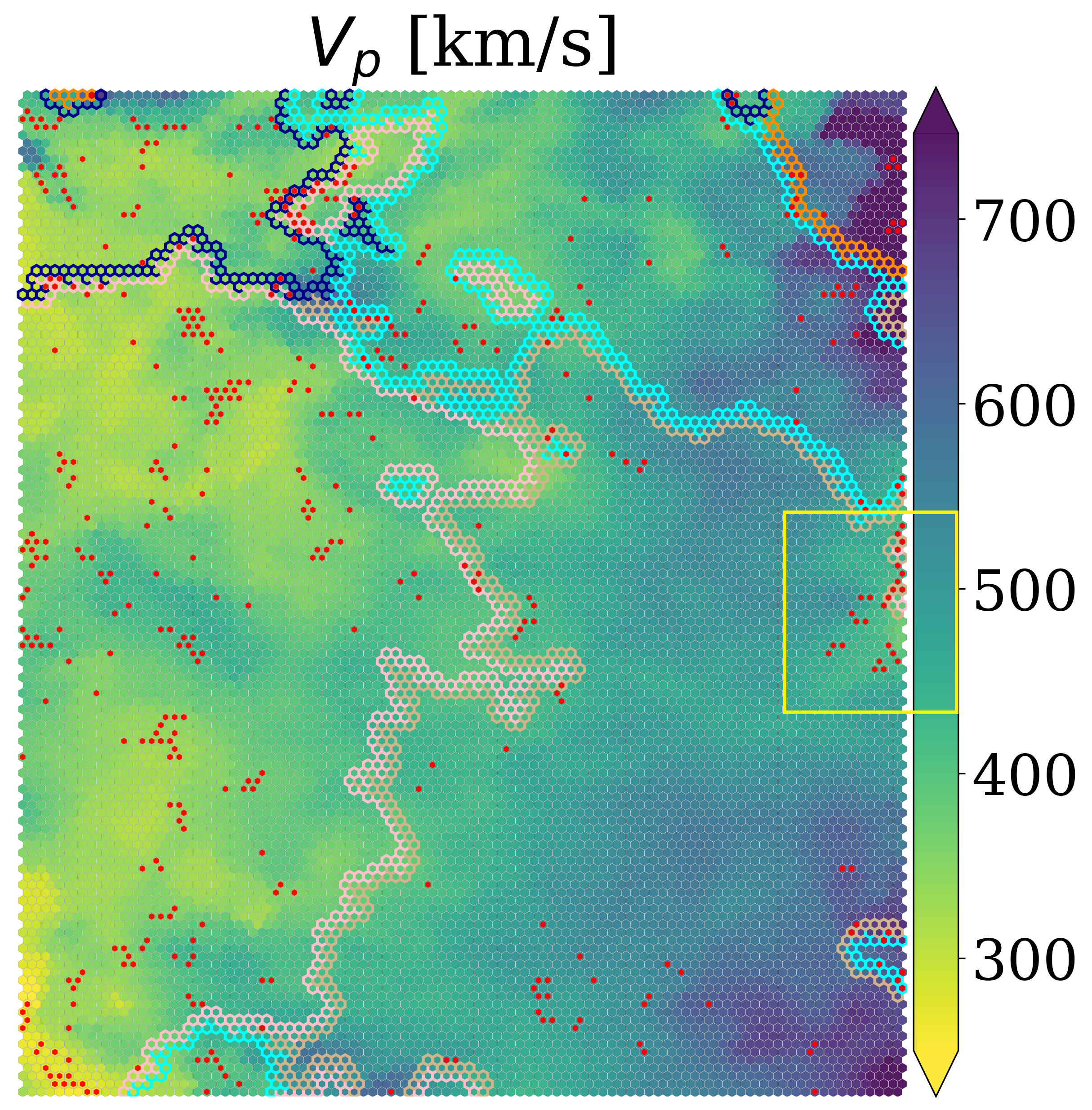}
	      \caption{Detailed picture of the velocity feature map, highlighting in yellow a region of the HAW cluster where the velocity is lower and some of the reconnection neurons are clustered.}
	      \label{fig:hexvhighlight}
      \end{figure}
      \begin{figure}[!hbt]
         \centering
         \includegraphics[width=\linewidth]{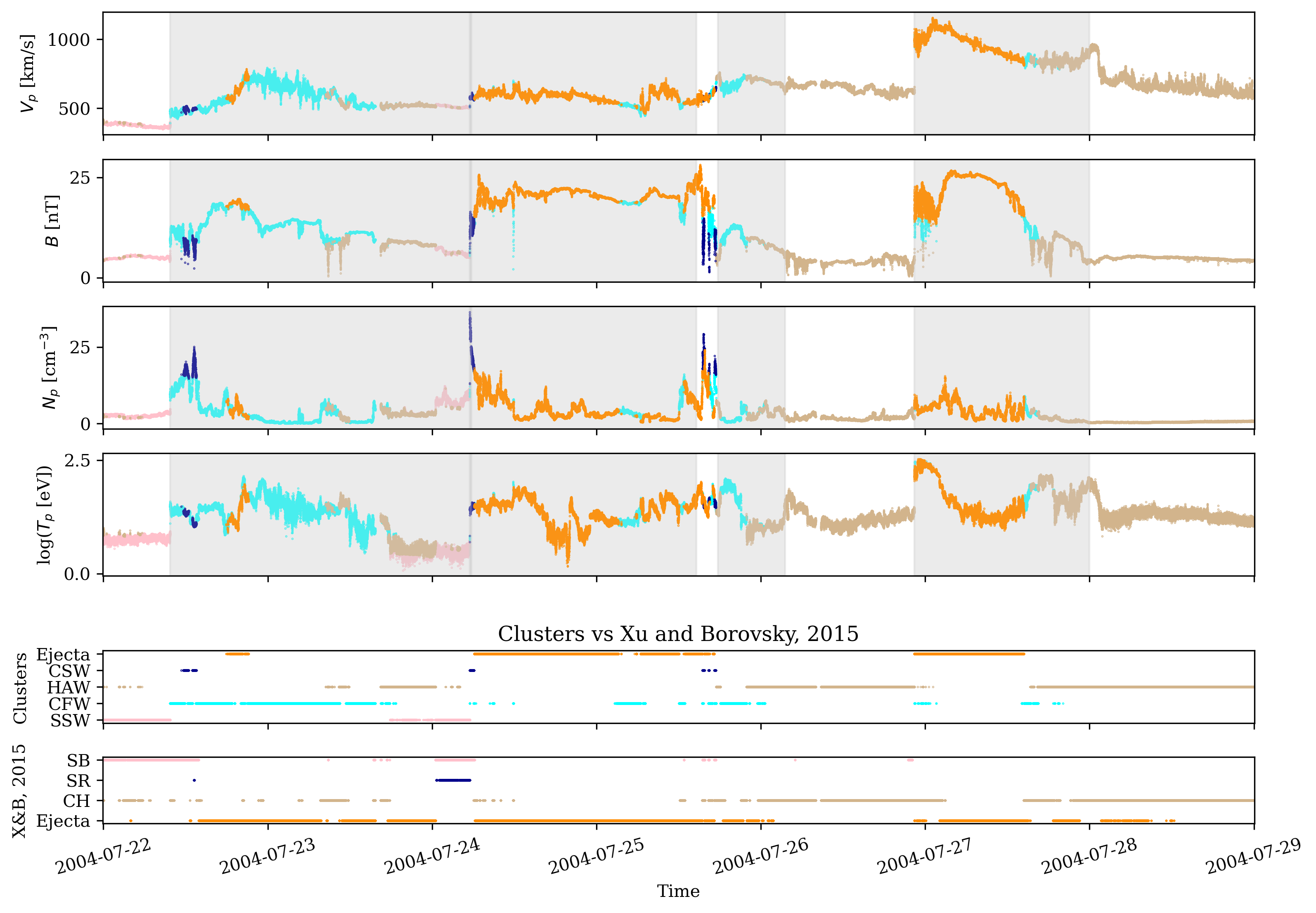}
         \caption{(\textit{Top}) Clustered time series for four ICMEs from \citet{nieves1} (in gray) during July 2004. It is noticeable that the orange cluster is associated with ejecta with a clear signature of a magnetic cloud, whereas the cyan cluster is associated with compressed fast wind and related to events where the magnetic cloud is not clearly defined. For the first three ICMEs, it is noticeable that we have a first phase where the slow wind is compressed (blue), followed by a fast compressed wind, and depending on the event, one has an orange cluster, which is associated with magnetic clouds. The exception is for the last ICME, where a strong shock occurred, and we have no compressed slow wind, but only compressed fast wind. (\textit{Bottom}) Comparison between \citet{Xu} classification. In this case, the Xu and Borovsky classification better matches the ICMEs.}
         \label{fig:icmesjuly}
      \end{figure}

\end{appendix}

\end{document}